\documentclass[twocolumn,apj]{emulateapj}
\usepackage{relsize,cancel, graphicx, dcolumn,graphicx,color,booktabs,bbold, microtype,braket,mathrsfs,mathtools,simplewick,tikz,empheq,amssymb,amsmath,bm}
\usepackage[colorlinks,plainpages=false,linkcolor=black,urlcolor=blue,citecolor=blue,pdfpagemode=UseNone,pdfstartview=FitH,allcolors=blue]
{hyperref}
\usepackage[T1]{fontenc}
\usepackage{cleveref}
\usepackage{amsmath}

\begin{document}
\title{Real-time flare prediction based on distinctions between flaring and non-flaring active region spectra}
\author{Brandon Panos\altaffilmark{1,2}, Lucia Kleint\altaffilmark{1,3}}\
\altaffiltext{1}{University of Applied Sciences and Arts Northwestern Switzerland, Bahnhofstrasse 6, 5210 Windisch, Switzerland}
\altaffiltext{2}{University of Geneva, 1205 Geneva, Switzerland}
\altaffiltext{3}{Leibniz-Institut f\"ur Sonnenphysik (KIS), Sch\"oneckstrasse 6, D-79104 Freiburg, Germany.}

\begin{abstract}
With machine learning entering into the awareness of the heliophysics community, solar flare prediction has become a topic of increased interest. Although machine learning models have advanced with each successive publication, the input data has remained largely fixed on magnetic features. Despite this increased model complexity, results seem to indicate that photospheric magnetic field data alone may not be a wholly sufficient source of data for flare prediction. For the first time we have extended the study of flare prediction to spectral data. In this work, we use Deep Neural Networks to monitor the changes of several features derived from the strong resonant Mg II h\&k lines observed by IRIS. The features in descending order of predictive capability are: The triplet emission at 2798.77 $\text{\AA}$, line core intensity, total continuum emission between the h\&k line cores, the k/h ratio, line-width, followed by several other line features such as asymmetry and line center. Regions that are about to flare generate spectra which are distinguishable from non-flaring active region spectra. Our algorithm can correctly identify pre-flare spectra approximately 35 minutes before the start of the flare, with an AUC of 86 \% and an accuracy, precision and recall of 80 \%. The accuracy and AUC monotonically increases to 90 \% and 97 \% respectively as we move closer in time to the start of the flare. Our study indicates that spectral data alone can lead to good predictive models and should be considered as an additional source of information alongside photospheric magnetograms.
\end{abstract}

\keywords{Sun: flares; chromosphere --- line: profiles  --- methods: data analysis; statistical}

\section{Introduction}\label{Introduction}
Solar flares are magnetically driven phenomena, whereby \textit{free magnetic energy} is slowly accumulated via subphotospheric motions which sweep coronal potential fields into successively higher energy configurations \citep{Adding_free_energy}. These stressed magnetic fields can undergo an impulsive reconfiguration known as \textit{magnetic reconnection}, into a lower energy state, dissipating roughly $10^{32}$ ergs of  magnetically stored energy into kinetic energy used to accelerate electrons and protons from coronal heights into the comparatively thicker atmosphere of the chromosphere \citep{Magnetic}.

Vector magnetograms derived from the polarized structure of magnetically sensitive spectral lines allow us to approximate the accumulation of magnetic free energy by measuring the angle between the transverse component of the field and the observed line of sight field. \cite{Shear} showed that this measure of  magnetic \textit{shear} was strongly coupled to flare activity. 

With the launch of  the Solar Dynamics Observatory (SDO) in 2010 \citep{SDO}, important magnetic features such as these could be monitored continuously by SDO's Helioseismic and Magnetic Imager (HMI),
improving upon previous measurements of the Solar and Heliospheric Observatory's Michelson Doppler Imager (MDI) \citep{MDI}. With HMIs unprecedented magnetic coverage, and a rising awareness of powerful machine learning techniques, flare prediction based on the evolution of magnetic features has become a topic of intense study.

Machine learning algorithms combine magnetic features to form complex non-linear functions known as \textit{models}. These models can then be used to predict with a certain probability, whether active regions will produce a flare within a given period of time (ordinarily 24 hrs) or not. A model's success is loosely based on the number of correct predictions it makes in this binary flare/no-flare classification problem, with a variety of metrics used to interpret different aspects of the models performance. The majority of these metrics are sensitive to the ratio of flare/no-flare observations within a particular data set. This provides an ambiguity, which makes it difficult to track progress in the field, especially for early publications. To address the problem of \textit{class imbalance}, \cite{TSS_sugest} suggested the use of a ratio invariant skill score known as the \textit{true skill statistic} (TSS), with a maximum perfect prediction score of 1, random guess score of 0.5 and an adverse score of -1. Most recent publications have adopted this metric, making it easier to compare research results. Although the TSS is ratio invariant, it is still vulnerable to some degree of subjectivity, since one can fine tune a model's parameters to maximize the TSS score at the expense of other metrics.\\

\cite{Leka_2003} pioneered the use of magnetic field data for flare prediction, using a linear classifier known as \textit{discriminant analysis} to rank the predictive value of a number of magnetic parameters collected by the University of Hawaii's Imaging Vector Magnetograph. They concluded that flare prediction requires the combination of several predictive variables, with no one variable being either necessary or sufficient for flare production. \cite{Leka_2007_Limited} extended this study to include more active regions, and achieved an accuracy of 92\% for large flares, however, because of the large class imbalance, predicting all active regions as non-flaring would result in a similar high score of 90\%. The authors therefore concluded that (at least for their parametrization), the state of the photospheric magnetic field had little to no bearing on whether an active region would flare or not.

\cite{MDI_2} used space based MDI line-of-sight magnetic maps, free from atmospheric distortion, in combination with a Neural Network to predict 150 large flares with an accuracy of 69\%, approaching that of an experienced human observer \citep{human}. \cite{Monica_HMI} performed an extensive study on 1.5 million HMI active region patches of vector magnetic field data and used a non-linear classifier known as a \textit{support vector machine} (SVM) to predict X- and M-class flares based on 25 magnetic features. They generated two data sets based on two distinct labeling criteria. The first criterion assigned active regions to the negative/non-flaring class if a flare did not occur within 24 hours from sample time, while the second labeling scheme required a flare free window of $\pm48$ hours for a negative class classification. They achieved TSS scores of 0.76 and 0.82 respectively. In addition to large flares, \cite{Random_Forest_HMI} included B and C flares in a multiclass prediction study using \textit{random forests} and 13 HMI magnetic features, achieving TSS scores of 0.70, 0.33, 0.50, and 0.29 for B-, C-, M-, and X-class flares respectively, and attributed the poor performance on X-class flares to the lack of sufficient big flare training data produced by solar cycle 24. \cite{FlareCast} used conventional statistical techniques as well as several machine learning algorithms including multi-layer perceptrons, SVMs, and random forests to predict flares based on 13 novel magnetic field features derived from HMI. They achieved TSS scores of 0.74 and 0.60 for flares in excess of M1.0- and C1.0-class respectively. The temporal evolution of the magnetic field was first folded into predictive models by \cite{ShortTermSF}, who used a \textit{decision tree classifier} and a \textit{learning vector quantization} to improved the \textit{precision} of Wang's 2008 results by 10\%. \cite{LSTM_HMI} took this one step further and used a state of the art recurrent neural network known as a \textit{long-short term memory network} to captured the temporal evolution of 25 HMI magnetic features as well as 15 flare history parameters, achieving TSS scores of 0.88, 0.79 and 0.61 for flares in excess of M5.0-, M1.0- and C1.0-class flares respectively, with the implication that larger flares are easier to predict.\\

Although the complexity of machine learning models has increased with each successive publication, the predictive performance has reached a plateau. Many studies based on HMI data report that a small fraction of their features contain most of the predictive power, and that many of the features may be redundant permutations of one another. To this end, researchers have began to experiment with alternative types of data sets. For instance, \cite{UV_Brightening} tested several machine learning models based on 60 features derived from the classical HMI magnetograms conjoined with flare history, soft X-ray and UV emission. They found that UV brightening ranked amongst the top five predictive features. With this lead, \cite{Monica_Corona_HMI} used HMI photospheric vector-magnetic field data in conjunction with AIA image data from several passbands covering the chromosphere, transition region, and corona in order to capture the elusive and not yet well understood traces of the flare \textit{triggering} mechanism. They achieved similar results to those in the literature, with TSS scores between .70 and .85 depending on the combination of used features, and concluded that the predictive capability of the data may be saturated.

An attempt to find possible pre-flare traces within the ultraviolet passbands requires a change of time scale. It is well known that the time scale for magnetic free energy accumulation is of the order of several days, however, many flare precursors occur only several minutes before flare onset, for instance: Activation of prominences from the motion of adjacent large scale magnetic structures perturbing the supporting magnetic field have been observed 15 minutes before major flare activity \citep[e.g.][]{Kleint_2015}. Slowly leaking magnetic energy from the rearrangement of unstable magnetic fields can lead to pre-flare heating and enhanced X-ray and UV levels several minutes before flare onset \citep{Slow_leak}, and spikes associated with microflares confined within the transition region with no X-ray counterpart have been found to litter the light curve of the \ion{Si}{4} line 30 to 60 minutes before the onset of some major flares \citep{Si_Microflare}.\\

Following the current spirit of the scientific community to integrate alternative sources of data, we investigate the utility of spectra for solar flare prediction. Our ambition for this paper is not to provide a predictive model for solar flares, but to examine the usefulness of solar spectra in distinguishing between pre-flare active regions and active regions that do not result in a flare. In particular, we look at spectral profiles associated with the once ionized magnesium (\ion{Mg}{2}) h\&k lines. Our study asks two basic questions: 1) Are spectra generated in pre-flare and non-flaring active regions distinguishable? 2) If so, does this distinction become more apparent closer to flare onset? To answer these questions, we attempt to disentangle the profiles from pre-flare and active regions by successively increasing the model complexity, beginning with a basic \textit{dimensionality reduction technique} and ending with the application of a \textit{deep neural network}. In addition to this, we also examine the separability of spectra from quiet Sun and sunspot regions.\\

The paper is organized as follows: In section \ref{Data}, we introduce the spectral data, how it was parameterized and collected. In section \ref{low-dimensional representations}, we visualize the high dimensional data using dimensionality reduction techniques. In section \ref{Supervised learning}, we introduce several supervised learning models and discuss the different metrics used to judge varying aspects of their performance. In section \ref{Space and time dependent performance}, we test the utility of our model as a flare prediction tool, and finally, in section \ref{Conclusion} and \ref{Outlook}, we conclude our research results and offer potential future avenues for the research to unfold.

\section{Data}\label{Data}
\begin{figure}[h] 
\includegraphics[width=.49\textwidth]{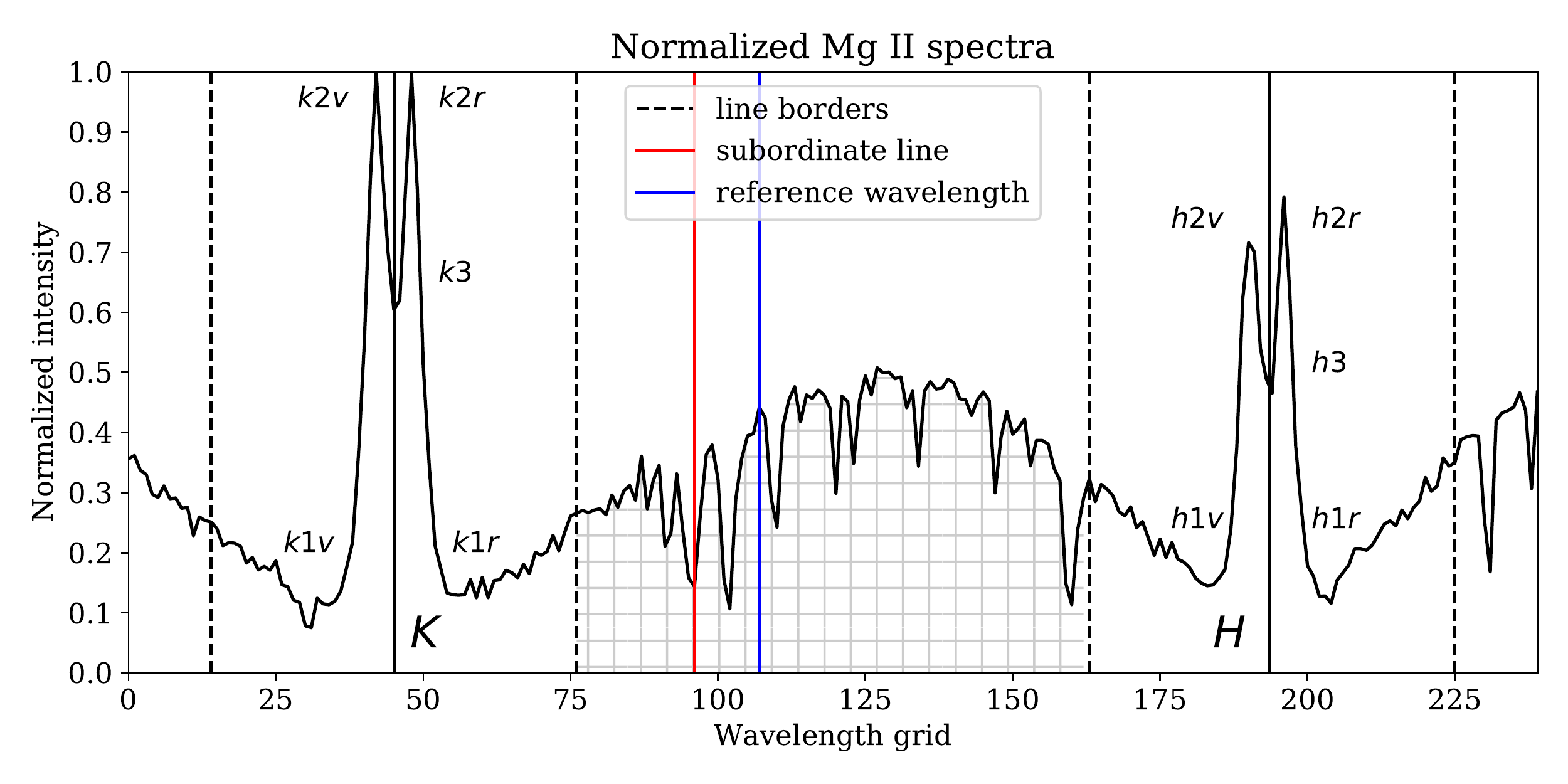}
\caption{The unhashed regions between the vertical dashed lines correspond to the k\&h-lines, while the hashed central region marks what we define as continuum emission, even though there are contributions from higher layers. The vertical red line indicates the position of the two blended red wing subordinate lines, which we define to be in emission when they surpass the intensity of the continuum at the location of the blue vertical line. The vertical dashed lines partition the k\&h-lines from the continuum.}
\label{profile}
\end{figure}

This study uses spectral data collected by NASA's Interface Region Imaging Spectrograph (IRIS) small explorer spacecraft \citep{IRIS}, which observes with high spectral resolution both the \ion{Mg}{2} h\&k lines seen in Figure \ref{profile}. The \ion{Mg}{2} lines have a formation height that extends over the entire chromosphere, with extensive wings following the temperature structure of the upper photosphere, 1v and 1r minimum emerging from the lower chromosphere, 2v and 2r peaks forming in the middle chromosphere and line core photons at k3 and h3 originating from an altitude just below the transition region. The h\&k resonant lines' relatively high opacity results in small geometrical thermalization lengths making them sensitive to temperature, density and velocity gradients \citep{sensitivity}. Additionally, the frequency dependent source function leads to a large variety of possible intensity profiles due to its complex interplay with the optical depth variation along the line of sight. The richness of line shapes coupled with a small thermal line width and formation height that extends over the entire chromosphere, makes the \ion{Mg}{2} profile an invaluable diagnostic tool for a large portion of the solar atmosphere.

\subsection{Feature selection} \label{Feature selection}
\begin{figure}[t] 
\includegraphics[width=.49\textwidth]{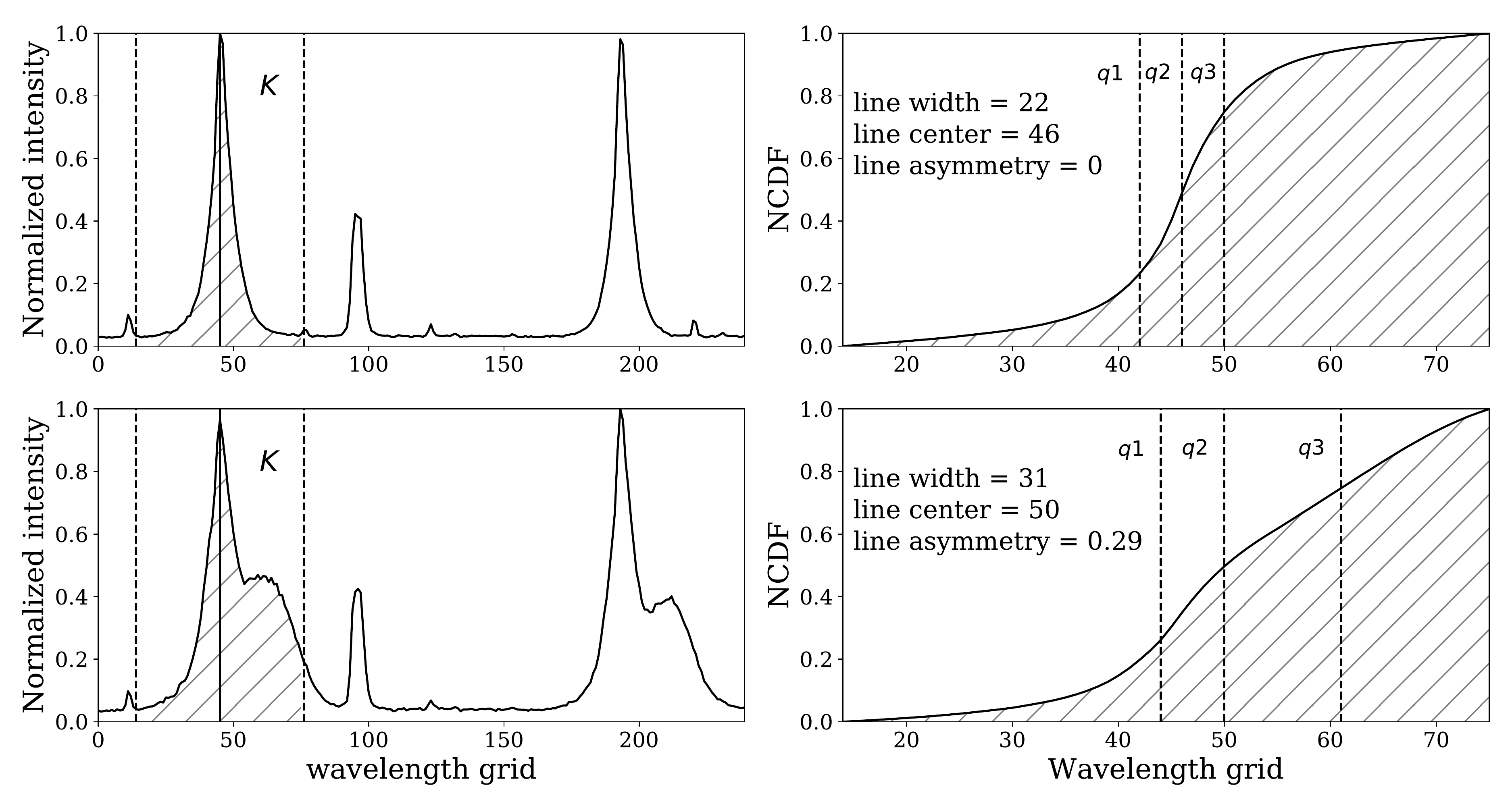}
\caption{Two normalized \ion{Mg}{2} profiles (left panels) and their corresponding cumulative distribution functions (right panels). Robust measures of line width, center and asymmetry can be derived from the three grid positions corresponding to the 25\% ($q1$), 50\% ($q2$), and 75\% ($q3$) quartiles, as indicated by the vertical dashed lines in the right panels. The NCDFs are constructed by creating a running sum of the hatched regions bounded by the vertical dashed lines in the left panels. The three features are only extracted for the k-line and their corresponding values can be seen in the right panels.}
\label{NCDF}
\end{figure}

Selecting an informative basis to represent the data, in this case spectra, is an integral part of machine learning. If we pass the entire profile to our algorithms, uninformative features such as the noisy small scale structure of the continuum will hamper the algorithm's ability to learn, since it first has to learn what information is useful and what information should be neglected. Although there are many deep learning models which automate this feature selection process, we find it more informative to describe each line profile in terms of a set of 10 parameters with known associations to observables. We list here the 10 selected features which provide us with a physical basis for describing each profile. Many of the listed features below (features, 1, 2, 3, 9 and 10), were shown by \cite{MgII} in a forward modeling paper with a 10-level-plus-continuum model atom, to be descriptive of temperature, velocity and density gradients within the chromosphere. To extract robust measures of the \textit{line center, line width and asymmetry}, we calculate each profile's normalized cumulative distribution function (NCDF) as illustrated in Figure \ref{NCDF}. All features except the continuum emission and k/h ratio are taken exclusively with respect to the k-line, since both lines are highly symmetric.

\begin{figure}[htb] 
\includegraphics[trim={4cm 2cm 4cm 3.5cm},clip, width=0.49\textwidth]{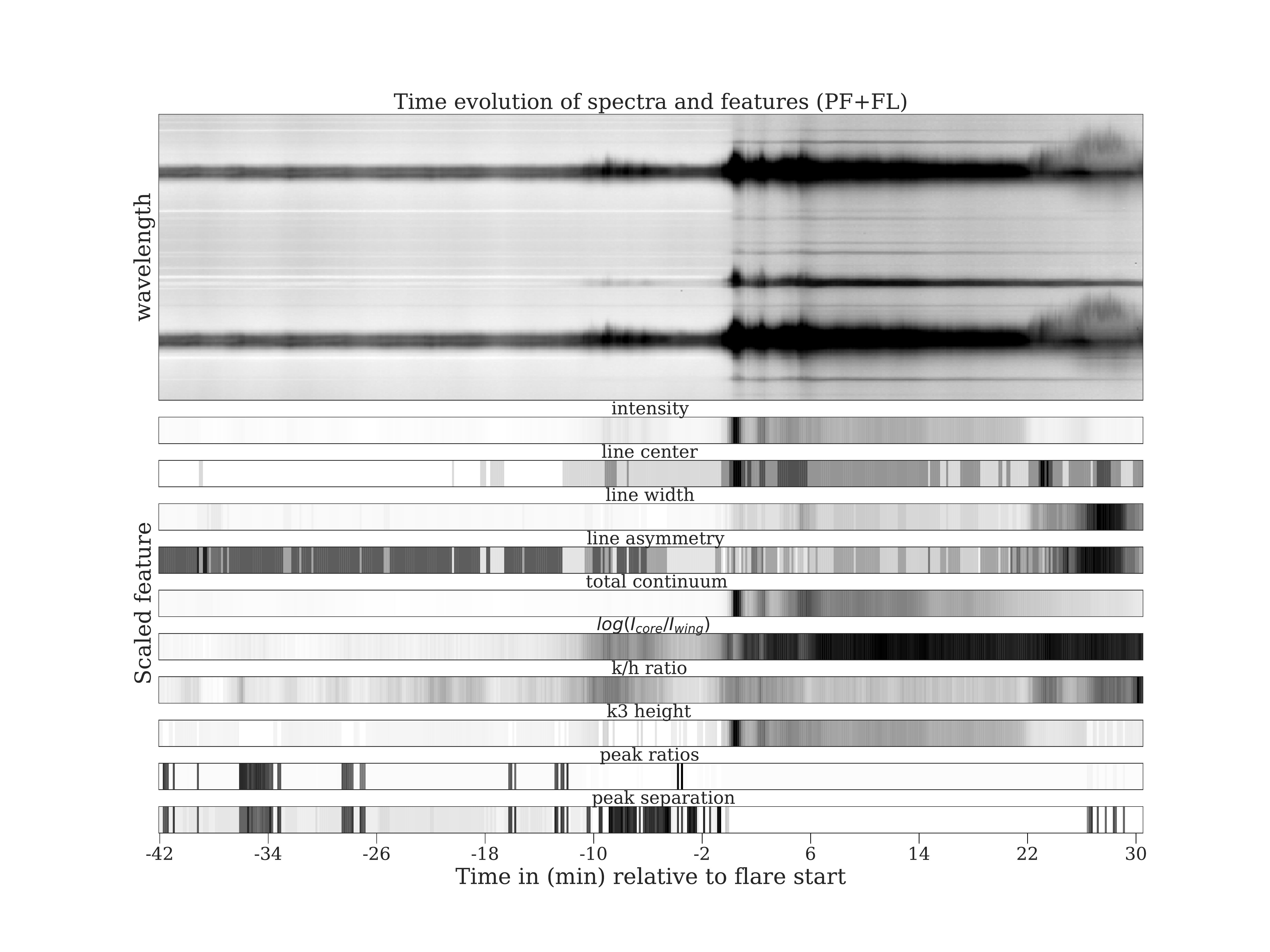} 
\caption{The upper panel shows a spectrogram of \ion{Mg}{2} leading up to and during an X1.6-class flare observed by IRIS on September 10, 2014 in pixel 400 along the slit. The 10 lower panels display the evolution of each feature. All features have been standardized to have zero-mean and unit-variance. The color code ranges from white to black, with white signifying lower, possibly negative values depending on the particular feature. }
\label{feature_evolution}
\end{figure}

\begin{enumerate}
\item \textit{Intensity}: For temperatures in excess of 6 kK, the peak intensity is strongly correlated to the gas temperature, with a correlation that starts to decrease for lower temperatures on account of the optical depth unity forming higher up in the chromosphere where the Planck and source function decouple. We divided each profile by its exposure time and selected the intensity as the maximum DN/s value.
\item \textit{Line center}: The doppler shift of the line core from its rest frame is an excellent diagnostic of the line of sight velocity in the upper chromosphere. This was selected as the 2nd quartile ($q2$) of each profile's cumulative distribution function (CDF). 
\item \textit{Line width}: Because of a small thermal width, line broadening is associated with non-thermal velocities. The line width was taken as $(q3-q1)$, where $q1$ and $q3$ are the 1st and 3rd quartiles respectively of each profile's CDF.
\item \textit{Line asymmetry}: Line asymmetry is an expression of non-thermal velocity flows and can take on values in the range $\pm1$, with positive and negative values being associated with up and downflows respectively. The line asymmetry expressed in terms of quartiles is given by $[ (q3-q2)-(q2-q1) ]/(q3-q1)$.
\item Total continuum: Since the local continuum radiation between the k\&h line cores is formed in LTE, the total intensity is dictated by the plasma temperature through the relation $I \simeq B \left[ T \left( \tau = 2 / 3 \right) \right]$, where $B$ is the Planck function and $\tau$ the optical depth. For solar flares, small scale heating events, and observations over sunspots, the continuum often appears flatter because of a different temperature structure. We define this feature as the sum of the hashed region in Figure \ref{profile} of a normalized profile. Therefore, larger total continuum values should be interpreted as intensity gains relative to the line core emission, and not taken in the absolute sense.
\item \textit{Triplet emission}:  The triplet emission on the red wing of the k-line core has been associated with lower atmospheric heating in the quiet Sun \citep{triplet_diagnostics}. To distinguish triplet emission from continuum emission we defined it as $\text{log}(I_\text{trip}/I_\text{wing})$, where $I_\text{trip}$ is the height at the red vertical line in Figure \ref{profile} and $I_\text{wing}$ the height of the profile at the blue vertical line. \cite{Stark} showed that the triplet lines at flare ribbons have a formation height similar to that of the h\&k cores, and therefore may not always be descriptive of the lower atmosphere. 
\item \textit{k/h ratio}: The k/h ratio can be used as a measure of opacity, with ratios of 1:1 indicating a more opaque atmosphere and ratios of 2:1 indicate optically thin line formation, see \cite{Panos}. We calculated the k/h ratio as the ratio of the integrated unhashed areas in Figure \ref{profile}.
\item \textit{k3-height}: \cite{Lucia} showed that the central reversal can go into emission if large densities or temperatures are introduced at the core formation height. In the former case, the increased densities recouple the Planck and source functions, while in the later case the temperature increase compensates for the decoupling allowing the source function to increase with height. Additionally, introducing layered velocity fields could fill in the reversal to create a single peak. The k3-height is taken to be the height of the k3 central minimum of a normalized profile. In the case of single peaked profiles, $\text{k}3=1$. 
\item \textit{Peak ratios}: The peak intensity ratios are correlated with upper chromospheric velocities, with large peak ratios indicating the existence of large velocity flows, and moderate ratios being linked to the difference between the average upper chromospheric velocity and the velocity at the peak formation height. We measure this as k2v/k2r. \cite{Bright_Gains} showed relative intensity gains in the violet peaks of \ion{Ca}{2} to be correlated with downwards moving atmospheres in a region just above the peak formation height, an attribute shared by the \ion{Mg}{2} line as shown by \cite{MgII}. Therefore, values in excess of one, correspond to larger 2v peaks, and consequently indicate downflows that shift the peak opacity to longer wavelengths. Conversely, values smaller than one are associated with upflows above the peak formation height.
\item \textit{Peak separation}: The separation in wavelength between the 2v and 2r peaks are well correlated with the difference between the minimum and maximum velocities in the region of atmosphere just above the peak formation height. We define this separation as the number of wavelength grid points between k2v and k2r.
\end{enumerate}
An evolution of these features leading up to and during an X1.6-class flare can be seen in Figure \ref{feature_evolution}. Before implementing any machine learning techniques, all features $x_i$ have been standardized by 
\begin{equation}
x_i^{\prime}=\frac{x_i-\overline{x}_i}{\sigma_i},
\end{equation}
where $\overline{x}_i$ is the mean, and $\sigma_i$ the standard deviation of said feature. This standardization allows the algorithms to converge faster and removes biases based on the specific ranges of each feature. 

\subsection{Data mining of classes}
This study is concerned with the distinguishability of spectra collected from different solar regions. The spectral data were therefore divided into four subclasses collected from 25 minute full field of view (FOV) observation windows over four solar regions: Quiet Sun (QS), sunspot (SS), non-flaring active regions (AR) and active regions resulting in flares, referred to as pre-flare regions (PF). The total number of instances analyzed in the conjoined data sets comes to 9 million spectral profiles, with the QS, AR, PF and SS data sets comprising 32 \%, 28 \%, 25 \% and 16 \% of the total data respectively. For the details of each observation we refer the reader to Tables \ref{qsssar} and \ref{pf}. Each spectrum was then labeled according to the region they were extracted from. For instance, all spectra from a SS observation carry a label of 0, while all PF spectra carry a label of 1. These four different labeled data sets are referred to as \textit{classes} in the standard machine learning vernacular. The PF 25 minute observation windows were positioned such that the end of each window coincided precisely with the 1 minute mark to flare onset, as defined by the Geostationary Operational Environmental Satellite (GOES). Only observations leading to X- or M-class flares were considered for the PF data set, while the AR data set is a composite of active regions that never resulted in a major flare within the full (unparsed) IRIS observation time window, which often exceeded 25 minutes. Ordinarily, the duration of each observation varies, with typical values ranging from a few minutes to several hours, depending on the data rates and several other limiting factors unique to the specific research goals. The small time interval of 25 minutes for each observation is a limitation imposed by the PF data set, as IRIS on average only observed 25 minutes before each major flare. As a consequence of using the full FOV, the data sets are expected to carry an intrinsic number of shared profiles. Additionally, the data sets are naturally coupled to some degree, since our selected active regions contain sunspots, and pre-flare regions ordinarily emanate from active regions, however, this does not preclude the possibility of these regions being remarkably distinct when observed in the near-ultraviolet.

\begin{figure}[htb] 
\centering
\includegraphics[width=0.5\textwidth]{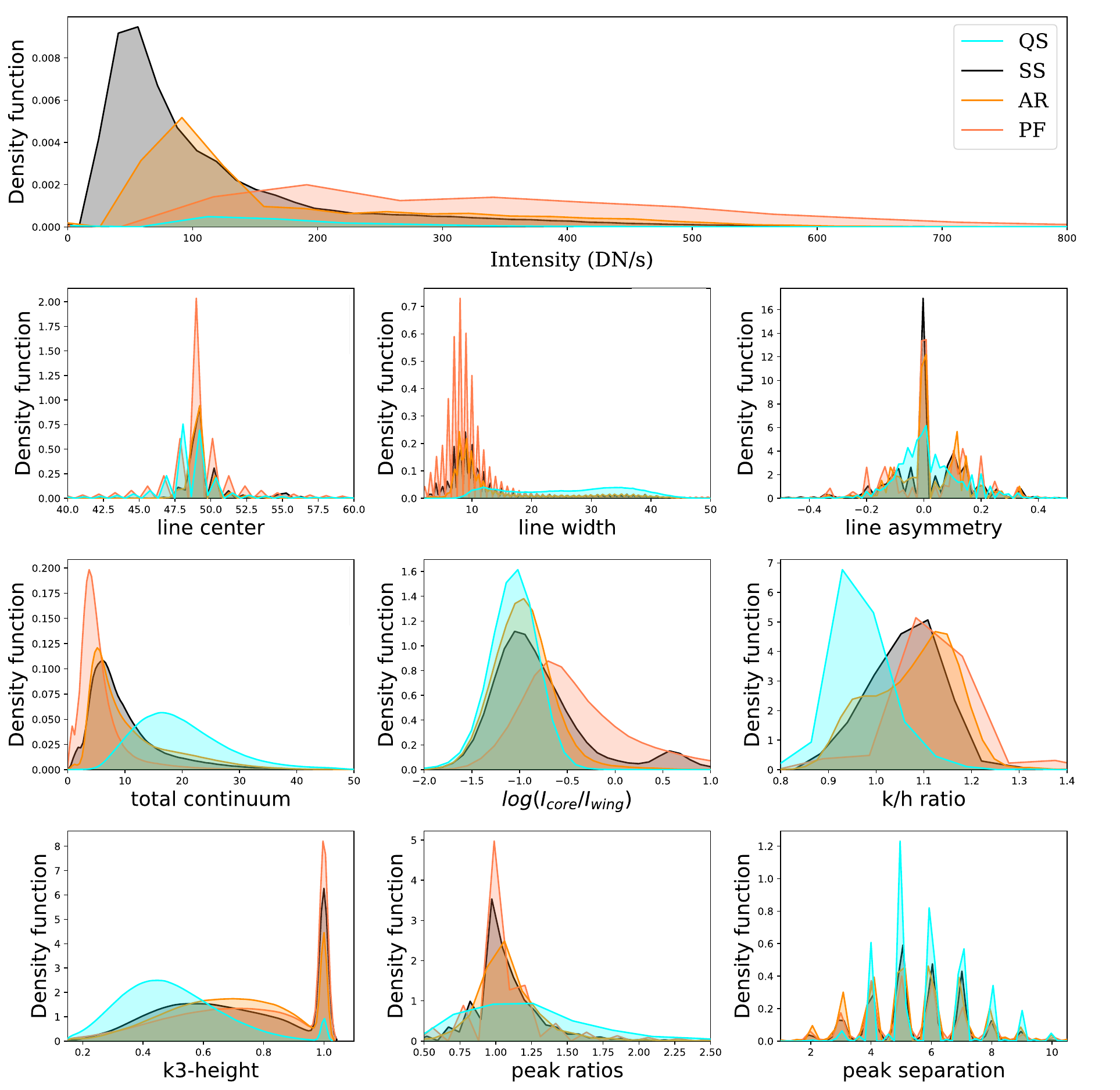} 
\caption{Spectral feature distributions for each of the four solar regions. Feature values are derived from normalized profiles on interpolated wavelength grids as shown in Figure \ref{profile}. Features such as line center, asymmetry, and peak ratio were calculated with respect to the k-line only.}
\label{feature_distributions}
\end{figure}

The IRIS observations were chosen in a way that minimized the possibility of introducing unwanted biases. If the observations from the four target regions are not evenly distributed in time and on the solar disk, machine learning algorithms could artificially distinguish between them based on some gross underlying temporal and geometric features not inherent to the physics of the Sun. For instance, if all  AR observations are collected at the limb and all PF observations at disk centre, then one might distinguish AR profiles from PF profiles based on the sum of the continuum emission, since center to limb variations result in a systematic decrease of the continuum (at least for plane parallel models). We note that this limb darkening effect was not directly observed by us, possibly because the contribution to the \ion{Mg}{2} line shapes from local atmospheric conditions far exceeds the effects introduced by the observation angle. Nevertheless, to avoid introducing "artifacts", we attempted to match the distributions of each regions observations in time and on the solar disk as closely as possible. Additionally, we ensured that AR and PF region activity was similarly represented by carefully monitoring SJI animations and keeping only those IRIS observations for which the GOES soft X-ray flux remained below the $10^{-5} \text{W}/\text{m}^2$ mark. The distributions of each of the four data sets in terms of the unstandardized 10 descriptive features can be seen in Figure \ref{feature_distributions}. The data has been fitted with \textit{kernel density estimators} (KDEs). A KDE can be viewed as the continuous counterpart of a histogram. For the case of a histogram, the height of each bar is determined by the number of points falling within the sharp boundaries of each bin. Histograms therefore do not make any assumptions about the underlying distribution of the data, and have a single free parameter (number of bins), which can only be used to degrade the resolution of the data. In contrast, a KDE uses a subtler weighting function with smooth edges, called a \textit{kernel}. The height or \textit{kernel density estimate} $\hat{f}_{h}$, at a particular point $x$ is determined by the number of contributing points falling within the kernel:
\begin{equation}
\widehat{f}_{h}(x)=\frac{1}{n h} \sum_{i=1}^{n} K\left(\frac{x-x_{i}}{h}\right),
\label{KDE}
\end{equation}
where $K$ is the kernel function (in our case Gaussian) and $h$ the so called bandwidth. The bandwidth determines the shape of the kernel function, and therefore affects how each point within the kernel is weighted. We have set $h=\sigma n^{-1/5}$, where $n$ is the number of data points for a specific region and $\sigma$, the variance for a particular feature of the data set. An alternative but equivalent interpretation is to imagine that kernels are centered at each data point and then summed together to produce a smooth continuous function. The denominator in Eq.(\ref{KDE}) divides out the area of the kernel, ensuring that the integral of the KDE equals unity. Note that the density function of a continuous variable at a single point need not be less than 1. Unlike histograms, KDEs do not degrade the data, but generate missing data in a controlled manner, and are therefore used to estimate probability density functions of unknown distributions.

\begin{figure}[htb] 
\centering
\includegraphics[width=.5\textwidth]{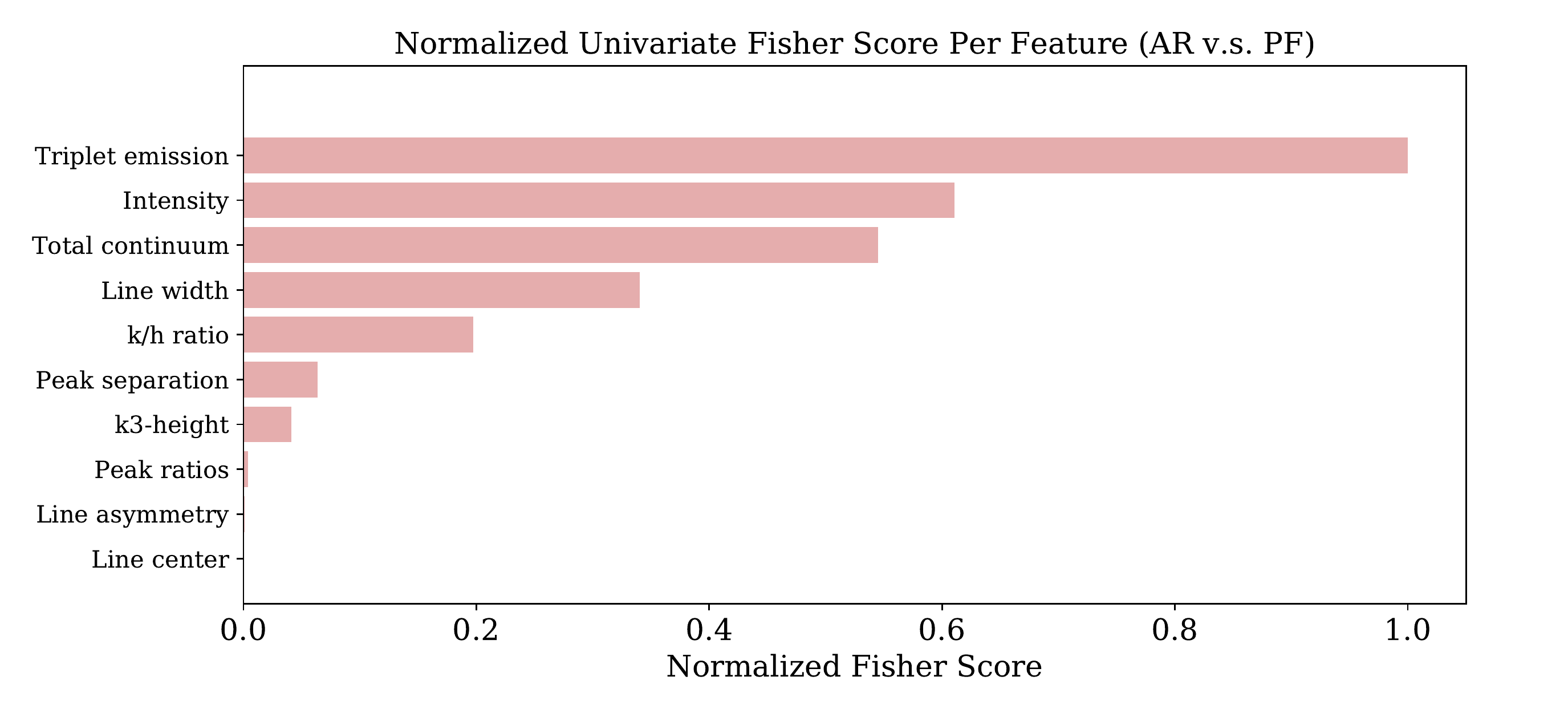}
\caption{F-scores for PF and AR feature distributions as seen in Figure \ref{feature_distributions}. The F-score tells us how dissimilar two distributions are. To a first approximation, features with higher F-scores have larger predictive capacities.}
\label{Fscores}
\end{figure}

The separation between each distribution in Figure \ref{feature_distributions} can be quantified with the use of a univariate F-score, or Fisher ranking score, which assumes uncoupled Gaussian distributed features. We have displayed the F-scores of each feature in Figure \ref{Fscores} for the case of PF/AR distributions. Higher F-scores imply larger dissimilarities between a feature's classes. In this first approximation, one can guess which feature is most or least important in distinguishing PF from AR spectra.

\section{low-dimensional representations}\label{low-dimensional representations}
In this section, we take spectra from all four regions, encode each profile in terms of our 10 features, and then use two different dimensionality reduction techniques: \textit{Principal component analysis} (PCA), and \textit{Stochastic neighborhood embedding} (t-SNE), to visualize the 10-dimensional data, by finding faithful low-dimensional representations referred to as \textit{embeddings}. Because t-SNE has a computational and memory complexity that is quadratic in the number of data points, modeling all 9 million spectra is not feasible. We therefore selected a subsample of 10,000 spectra from each region, as opposed to the 50,000 points used for the PCA plots. This subsampling adversely affects the representation of the underlying manifold. In addition to subsampling, all features were scaled by $x_i^{\prime}=(x_i-\overline{x}_i)/\sigma_i$ to avoid placing unwarranted precedence on features such as intensity, which cover a range of 0-800 as opposed to the subtler range of 0.8-1.4 covered by the k/h ratios. 

\begin{figure}[htb] 
\centering
\includegraphics[trim={0cm 0cm 0cm 0cm},clip, width=0.5\textwidth]{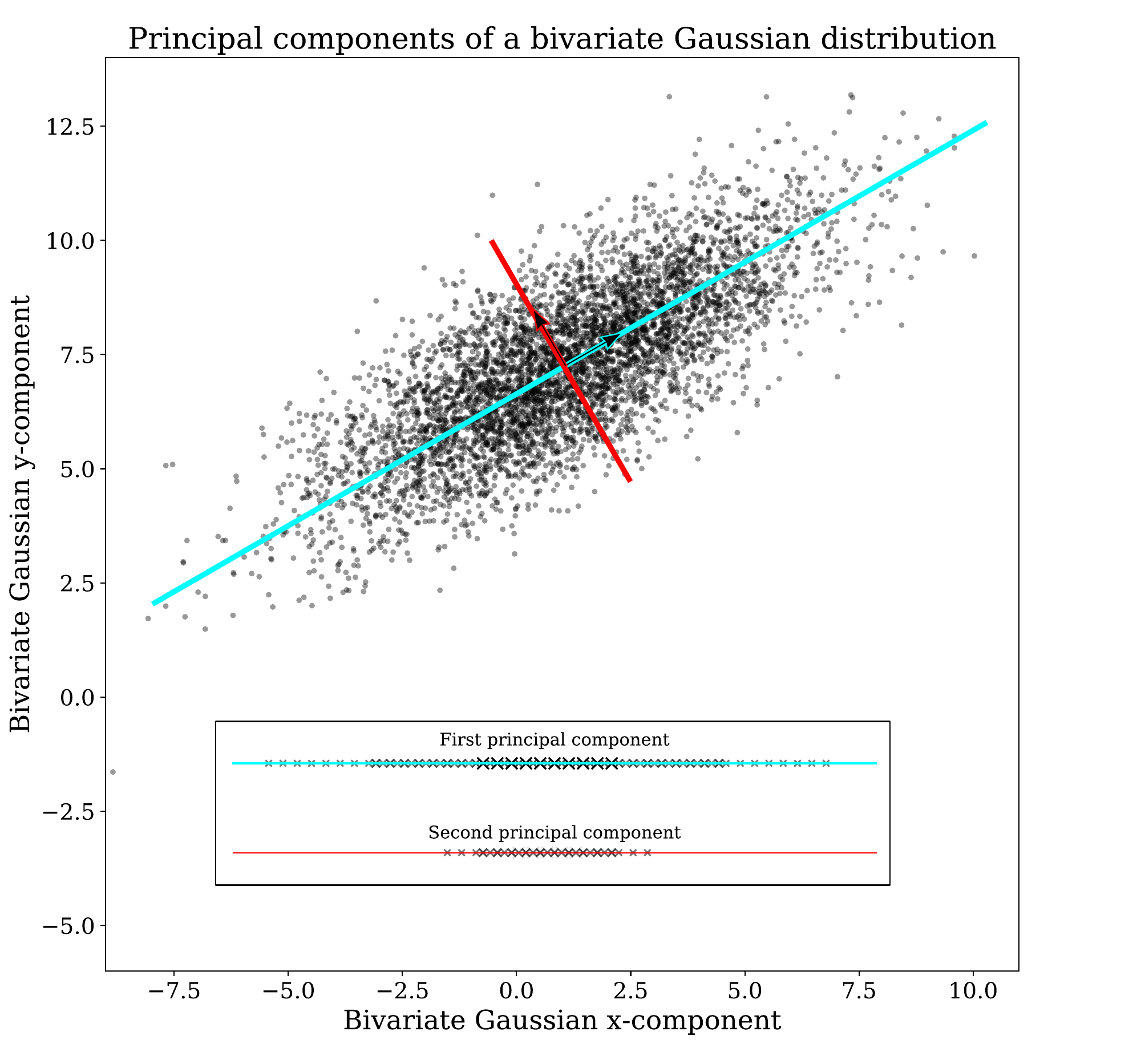} 
\caption{Projection of data from a bivariate Gaussian distribution onto the first two principal components. The first eigenvector of the covariant matrix (cyan) has a larger eigenvalue than the second component (red) and therefore captures more of the original data's variance.}
\label{toy_pca}
\end{figure}

\begin{figure*}[htb]
\centering
    \includegraphics[trim={0cm 0cm 0cm 0cm},clip, width=0.246\textwidth]{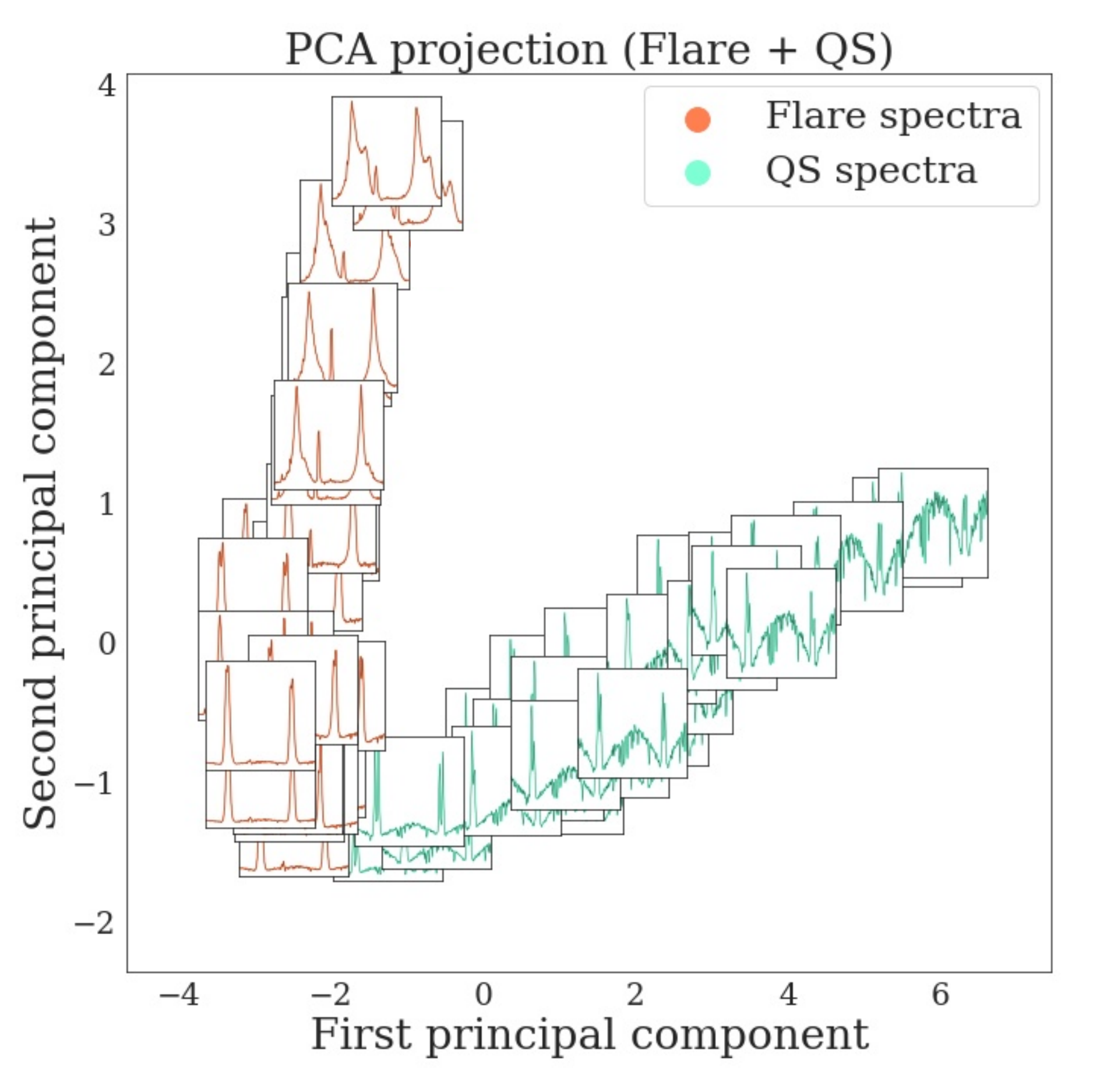}
    \includegraphics[trim={0cm 0cm 0cm 0cm},clip, width=0.246\textwidth]{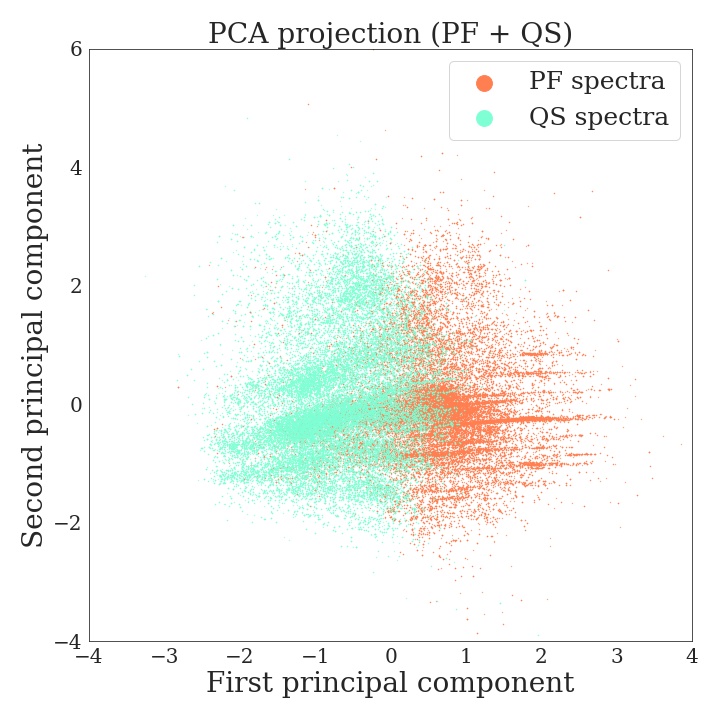}
    \includegraphics[trim={0cm 0cm 0cm 0cm},clip, width=0.246\textwidth]{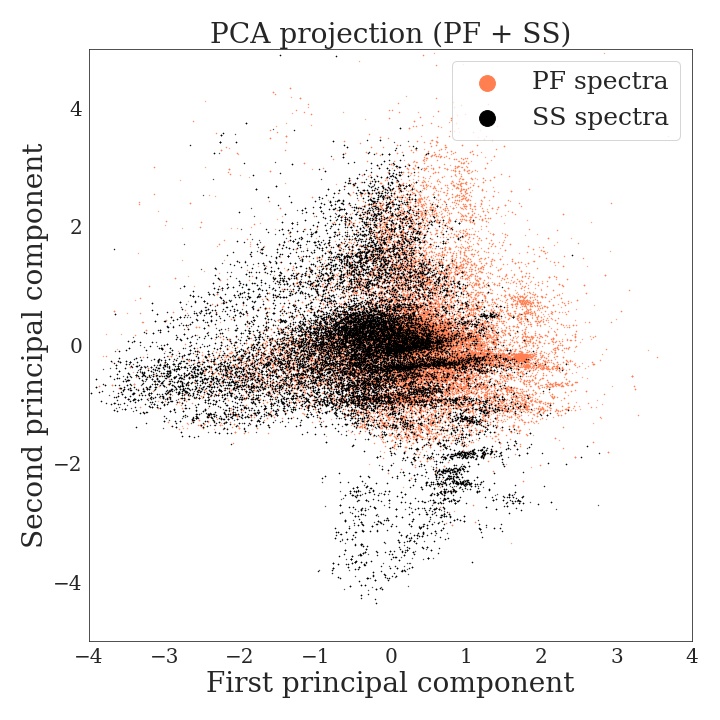}
    \includegraphics[trim={0cm 0cm 0cm 0cm},clip, width=0.246\textwidth]{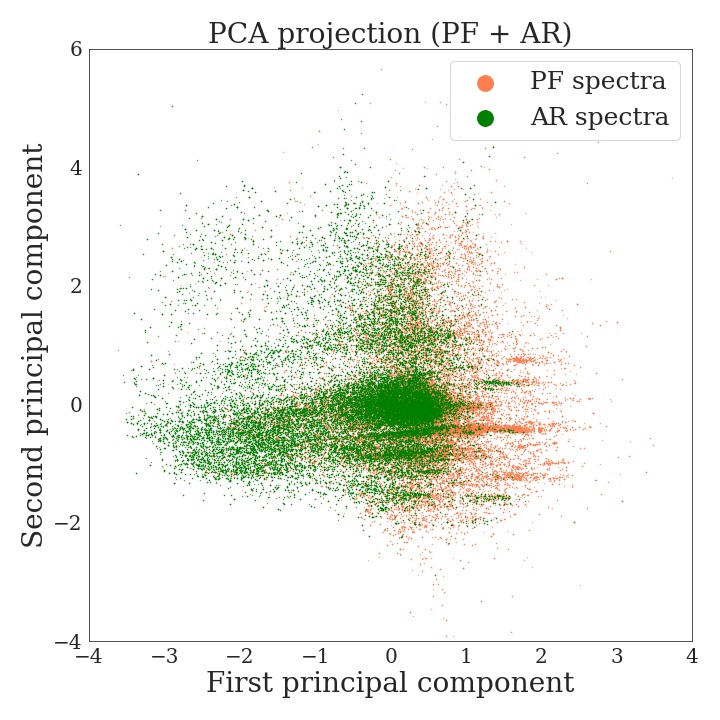} 
\caption{The left panel shows a PCA projection of quiet Sun (cyan) vs flare (orange) profiles. For clarity, we have replaced the data points by the profiles that they represent. The remaining panels show projections of 50,000 PF profiles (orange) along with profiles from the other three data sets. Note that before performing the dimensionality reduction, all profiles were converted into the 10-dimensional feature space and standardized.}
\label{pca_plots}
\end{figure*}

\begin{figure}[b] 
\centering
 \includegraphics[trim={0cm 0cm 0cm 0cm},clip, width=0.49\textwidth]{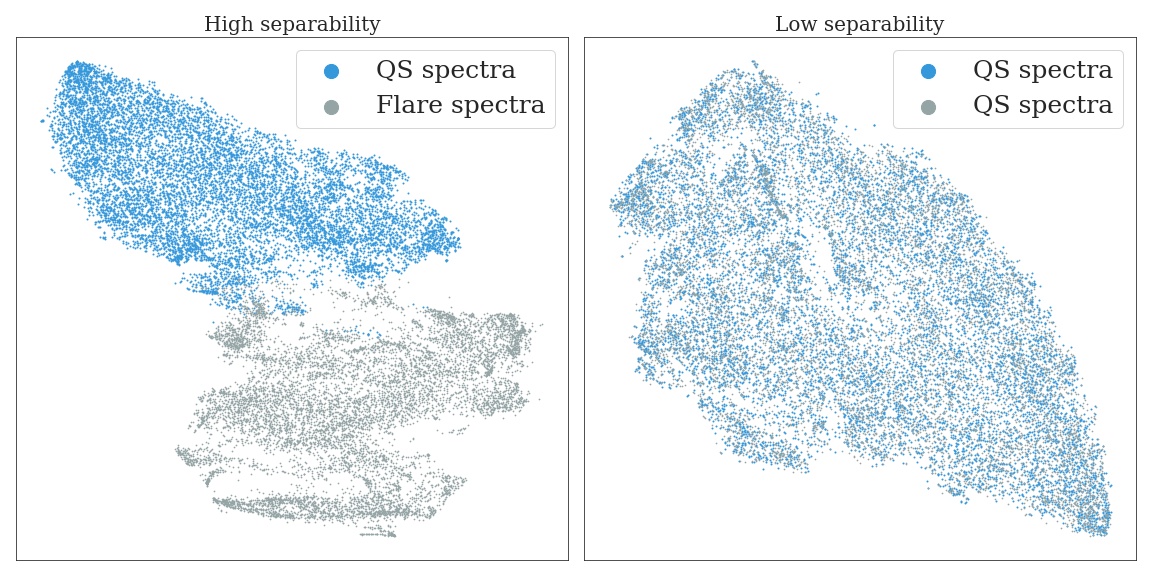}
\caption{t-SNE plot of flare vs QS spectra (left panel) and QS vs QS spectra (right panel). Because spectra from flares are vastly different from quiet Sun spectra, the left panel indicates what a t-SNE embedding would look like for two classes that are highly separable. In contrast, the panel on the right shows a t-SNE embedding of two data sets that are maximally inseparable. Separability should therefore be measured as the degree to which the classes mix, with cleaner, less confused outputs being associated with higher degrees of separability.}
\label{qsqs}
\end{figure}

\begin{figure*}[htb]
    \includegraphics[trim={1cm 1cm 1cm 1cm},clip, width=0.197\textwidth]{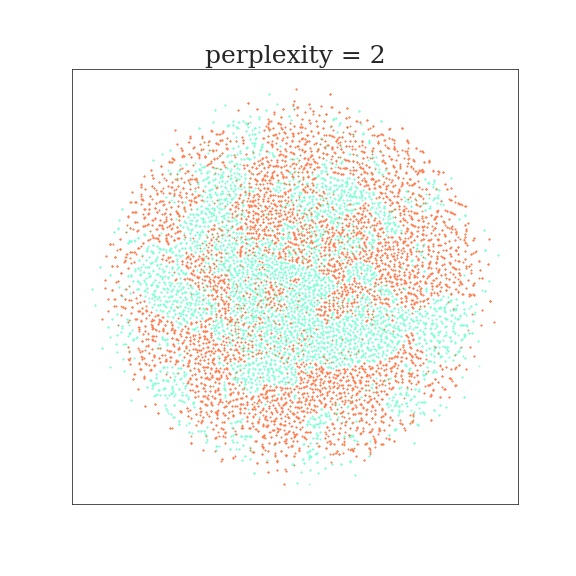}
    \includegraphics[trim={1cm 1cm 1cm 1cm},clip, width=0.197\textwidth]{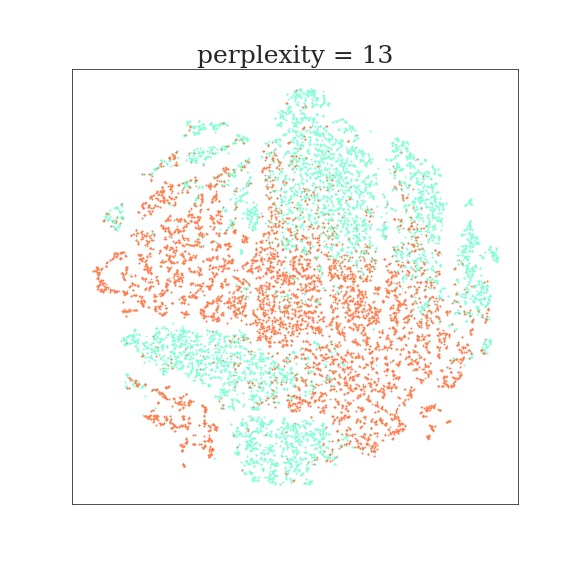}
    \includegraphics[trim={1cm 1cm 1cm 1cm},clip, width=0.197\textwidth]{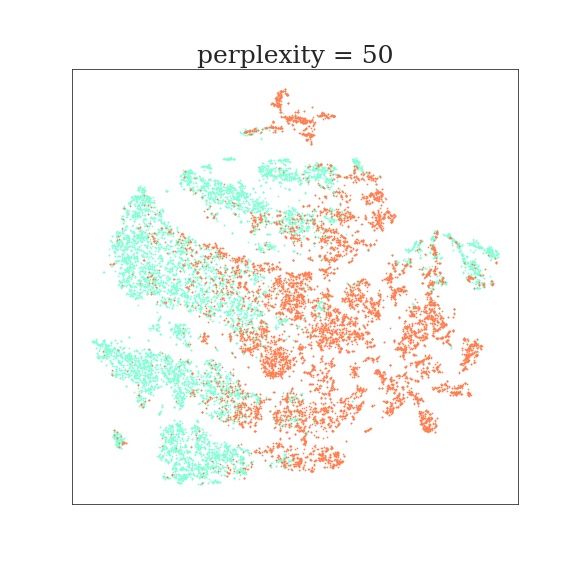} 
    \includegraphics[trim={1cm 1cm 1cm 1cm},clip, width=0.197\textwidth]{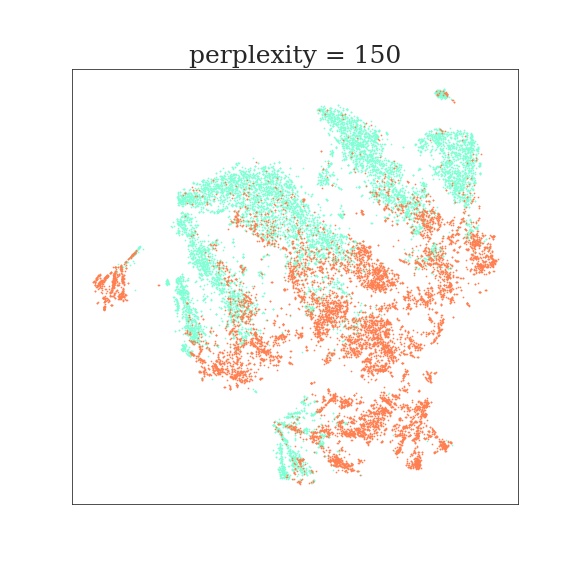}
    \includegraphics[trim={1cm 1cm 1cm 1cm},clip, width=0.197\textwidth]{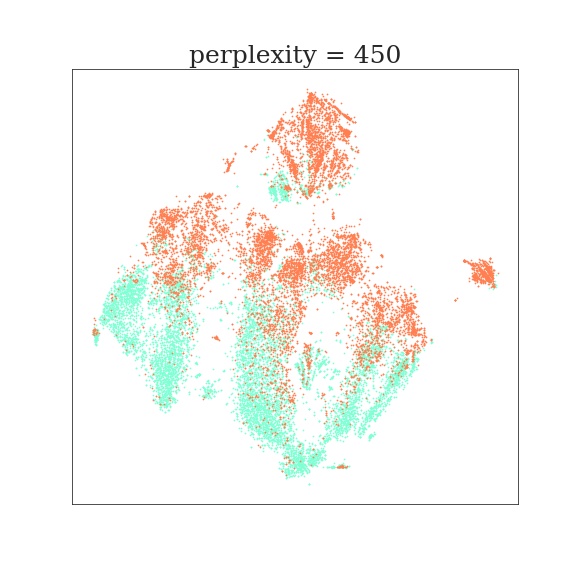} 
    
    \includegraphics[trim={1cm 1cm 1cm 1cm},clip, width=0.197\textwidth]{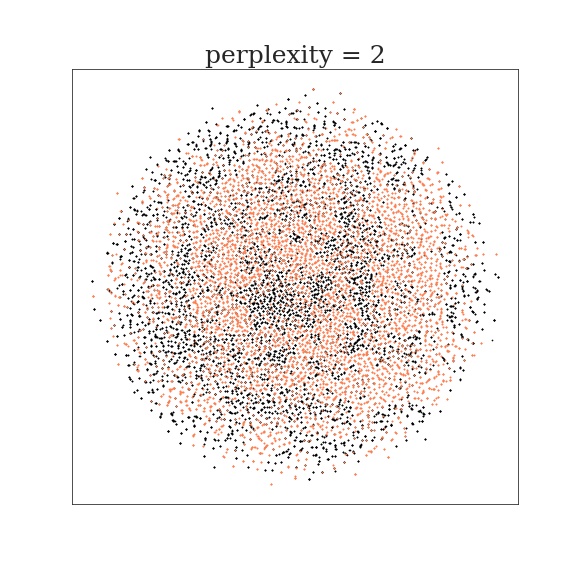} 
    \includegraphics[trim={1cm 1cm 1cm 1cm},clip, width=0.197\textwidth]{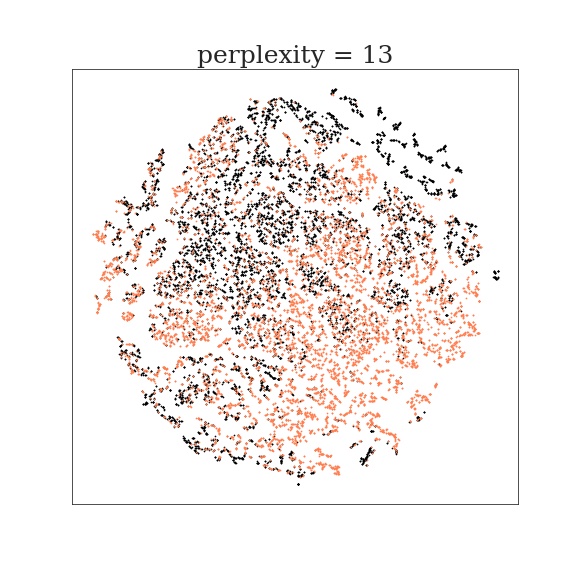}
    \includegraphics[trim={1cm 1cm 1cm 1cm},clip, width=0.197\textwidth]{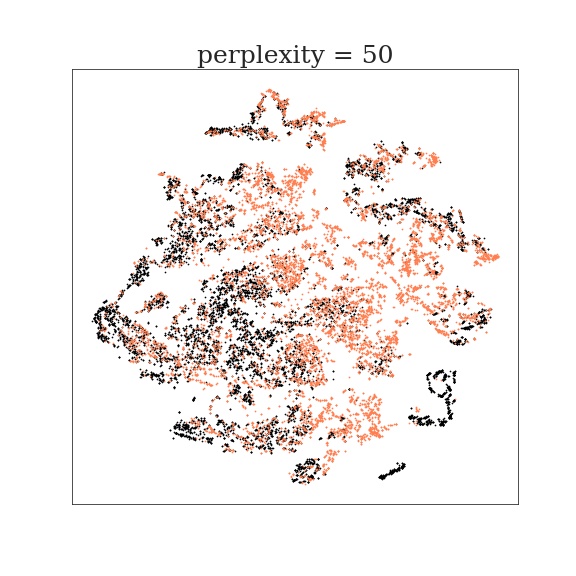}
    \includegraphics[trim={1cm 1cm 1cm 1cm},clip, width=0.197\textwidth]{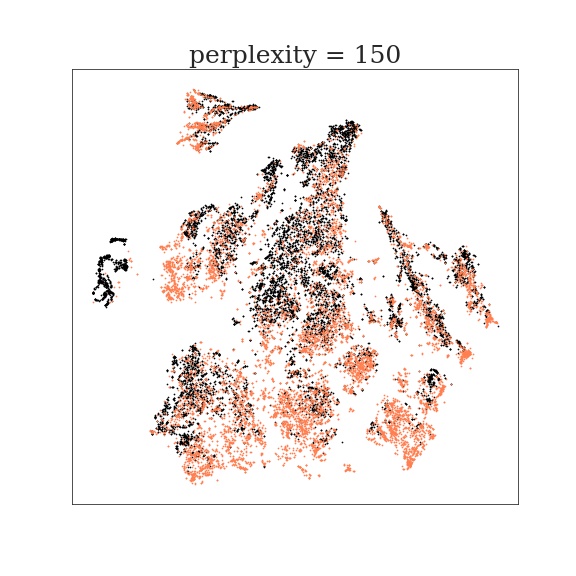}
    \includegraphics[trim={1cm 1cm 1cm 1cm},clip, width=0.197\textwidth]{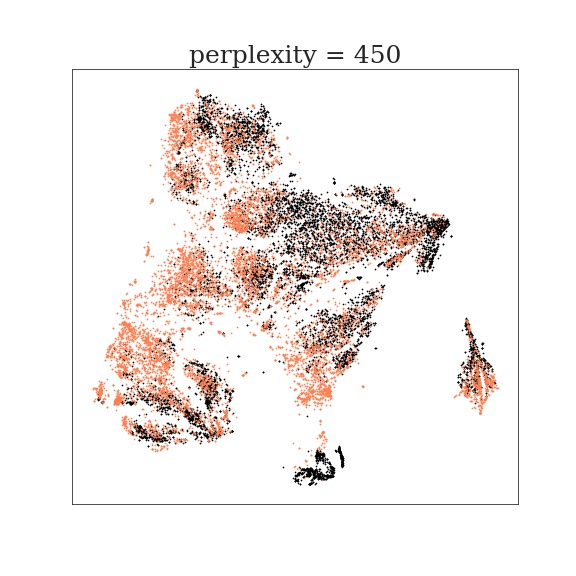} 
   
    \includegraphics[trim={1cm 1cm 1cm 1cm},clip, width=0.197\textwidth]{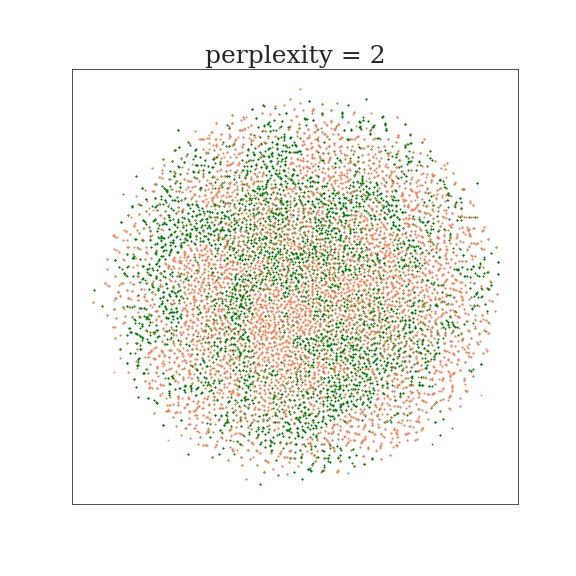}
    \includegraphics[trim={1cm 1cm 1cm 1cm},clip, width=0.197\textwidth]{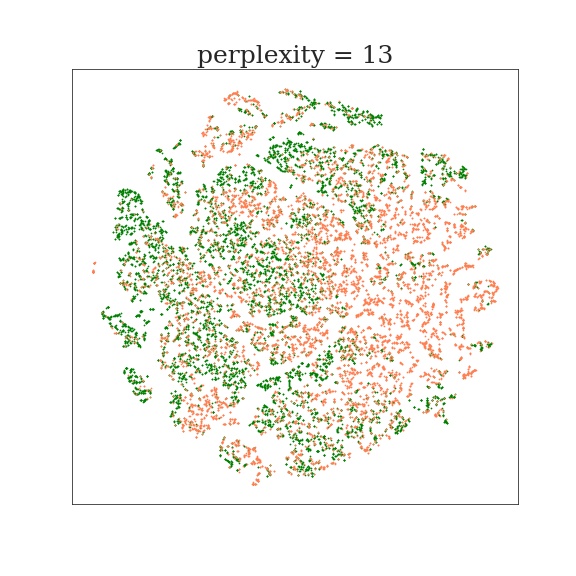}
    \includegraphics[trim={1cm 1cm 1cm 1cm},clip, width=0.197\textwidth]{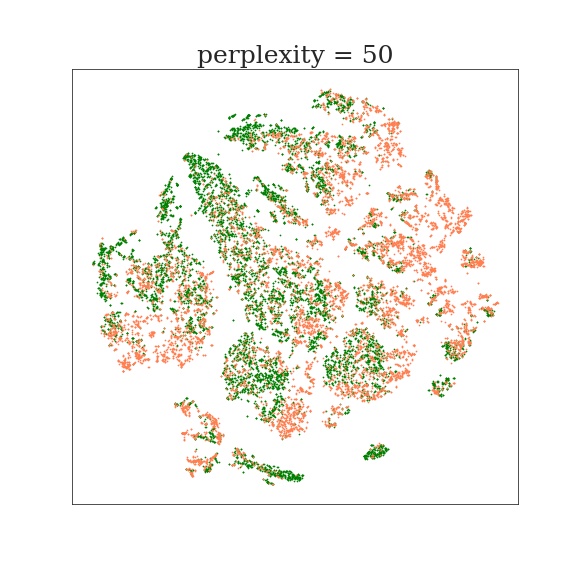}
    \includegraphics[trim={1cm 1cm 1cm 1cm},clip, width=0.197\textwidth]{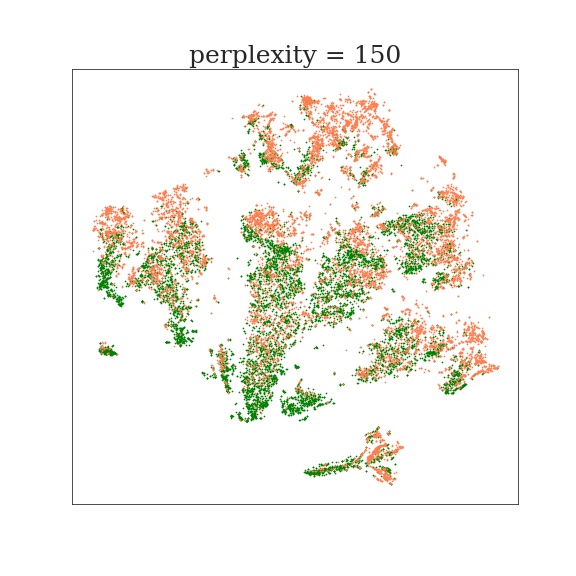}
    \includegraphics[trim={1cm 1cm 1cm 1cm},clip, width=0.197\textwidth]{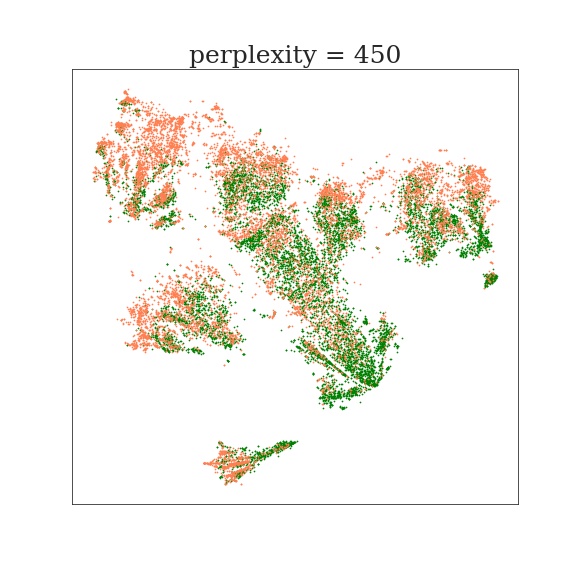} 
\caption{t-SNE plots of spectral data. Complexity increases from left to right, with each row comparing PF profiles (orange) to QS (cyan), SS (black) and AR profiles (green). Lower perplexities reflect the local structure of the data, while higher perplexities include more global features. It is custom to leave the axes of t-SNE plots blank since they have no physical interpretation. The PF profiles have been plotted first with the other profile types overplotted so that the degree to which the PF profiles are eclipsed can be used as an indicator of separability. The plots indicate that QS profiles are easily distinguishable from PF profiles, whereas SS and especially AR spectra are less distinct, however, profiles from these regions appear on average to occupy distinct subspaces.}
\label{t-SNE plots}
\end{figure*}

\subsection{PCA}
In the PCA scheme \citep{PCA}, high-dimensional data is projected onto a surface that best preserves the spread or variance of the data. As way of explanation, Figure \ref{toy_pca} shows a bivariate Gaussian distribution with points living in a 2-dimensional space. 1-dimensional representations can be obtained by projecting the data orthogonally onto lines defined by unit vectors. Two such projections are shown in the inset at the bottom of Figure \ref{toy_pca}. The low-dimensional representation resulting from projecting the data onto the cyan line is superior to that of the red representation, since it captures more of the data's variance and is therefore more descriptive of the original data set. The eigenvectors of the covariance matrix, commonly referred to as \textit{principal components}, form a convenient linearly independent basis for describing the data's variance, and eigenvectors with larger eigenvalues describe more of this variance. Instead of projecting our data onto a line, we constructed surfaces from the first two eigenvectors of the correlation matrix derived from the conjoined data sets (PF + QS), (PF + SS) and (PF + AR) separately. We then projected the spectra from 10-dimensions into 2-dimensions as seen in Figure \ref{pca_plots}, and compared the sample distributions of QS, SS and AR spectra with spectra derived from the PF data set (orange). In all three cases, the first two principal components captured roughly 54\% of the total variation within the data. The first panel on the left is for illustrative purposes, with the abstract data points being replaced by the spectra that they represent. We used QS spectra (cyan) and spectra from a flare (orange). Note that the two principal components appear to be descriptive of continuum emission and red wing enhancements, with large downflow profiles being found in the upper extremities of the plot. This clean correlation with single physical attributes is in no way guaranteed with PCA, and often the basis is composed of a complicated mixture of features. All remaining low-dimensional plots show only the abstract data points, since generating plots with 20,000 plus subplotted spectra is not feasible.

The plots indicate that PF and QS spectra are clearly different, however, the distinction between the first two principal components of PF and AR spectra is less obvious, with a large degree of overlap between their distributions. This does not mean that the PF and AR data sets are inseparable from one another, it may simply be that the PCA technique is not sophisticated enough to visualize a distinction that exists.

\subsection{t-SNE}
In contrast to PCA, t-SNE is a non-linear map that adapts to the underlying structure of the data \citep{t-SNE}. For a mathematical description of the algorithm, we refer the reader to the appendix. What sets t-SNE apart from other dimensionality reduction algorithms is its ability to probe data at different length scales. A parameter called the \textit{perplexity} $\alpha$, balances the importance between local and global structure representation. For instance, a large $\alpha$ penalizes the algorithm for representing the relationship between distant points poorly, whereas a small $\alpha$ only incurs a cost when nearby points are incorrectly represented. t-SNE is a complex dimensionality reduction algorithm that can lead to some unintuitive results. To better understand the output from t-SNE, we refer the reader to the excellent real-time simulations of \cite{Distil}. The separation of two classes can be judged with reference to Figure \ref{qsqs}. The left panel shows a t-SNE embedding of flare and QS spectra, in contrast to the embedding on the right, based on spectra from the same data set. Spectra generated in solar flares are dramatically different in shape than those generated in the quiet Sun, as a result, the left panel shows an embedding between classes that are highly separable. Analogously, the right hand panel shows a t-SNE embedding between classes that are maximally inseparable, since they come from the same data set. Unlike many dimensionality reduction techniques, the distance between classes (blue and grey data points), is a poor measure of separability, and in the t-SNE framework, separability is measured by the degree to which the classes are mixed, with cleaner unmixed groups indicating higher separability between classes. The reader can judge the degree of separation between two classes with reference to these baselines. To gain insight into the topology of the data set, multiple runs at a variety of perplexity settings are required. We took the same combination of data sets used to generate Figure \ref{pca_plots}, and constructed low-dimensional representations at a variety of perplexity settings. The results complement that of PCA, where one can clearly separate QS from PF profiles, with a distinction that becomes more faded for SS and especially AR profiles. It is important to note that the sense of distance between clusters as well as relative cluster size lose all meaning in the t-SNE framework. Topological concepts such as \textit{containment} however are preserved. Additionally, the algorithm is not guaranteed to converge to the same results, with multiple t-SNE runs at a fixed perplexity leading to different representations. Although different runs produce slightly different representations, we found no additional insight between these differences, and therefore only display the results from a single t-SNE run in Figure \ref{t-SNE plots}.

The magnitude of each profile's features have been plotted in Figure \ref{feature_bars} for the (AR+PF) data set at a perplexity setting of 450. Each panel corresponds to one of the 10 standardized descriptive features, with the upper left panel indicating which points belong to PF (orange) and AR (green) spectra. The figures indicate that triplet emission is the most descriptive feature when trying to separate AR from PF spectra.

A technique called \textit{Uniform Manifold Approximation and Projection} (UMAP), introduced by \cite{UMAP}, is a scalable non-linear manifold learning technique which uses concepts from Riemannian geometry and algebraic topology to produces results competitive to those of t-SNE, while arguably retaining more of the global structure. However, we found little discrepancy between the two methods on our data set, and preferenced the t-SNE algorithm based on its simpler and more comprehendible mathematical formalism.

\section{Supervised learning}\label{Supervised learning}
So far, we have taken \ion{Mg}{2} spectra from four different solar regions and visualized the high-dimensional data by finding faithful 2-dimensional embeddings, with the hope of observing distinct clusters corresponding to each region. Even if the data naturally separate into four subgroups, a low-dimensional embedding with the limited representative resources of just two-degrees of freedom, may not be capable of adequately displaying this separation. Additionally, all our techniques have used a single data set $\mathbf{X}$, containing all of the spectra as input. Although each profile $x^i$ (we use superscripts to indicate examples and subscripts to indicate features) has had an accompanying label $\text{y}^i$, indicating from which region (QS, SS, AR, PF) the spectrum originated from, this label has only been used to color code the results, and therefore has had no bearing on the process. We now transition to a set of more powerful machine learning methods that utilize both the labels actively within the algorithm, as well as the full 10-dimensional space.

\subsection{Neural networks}
Neural networks are what we call supervised learning techniques that require a labeled data set $\{\mathbf{X},\text{y}\}$. In our case, $x^i$ of $\mathbf{X}$ is a single \ion{Mg}{2} spectrum, with a corresponding label $\text{y}^i$ of $y$  indicating which solar region the spectrum originated from. From now on, we are only concerned with distinguishing PF spectra from non-PF spectra, therefore, our problem becomes a binary classification problem where $\text{y}^i$ can take on the value 0 or 1. The goal of supervised machine learning is to use a fraction of the labeled data, called the training set $\{\mathbf{X}_\text{train}, \text{y}_\text{train}\}$, to learn what is referred to as a \textit{model}. This model can then be used to predict whether a profile is a PF profile 1 or not 0, from a never seen before test set $\{\mathbf{X}_\text{test}, \text{y}_\text{test}\}$. The model that the algorithm learns is analogous to the human idea of a concept. In much the same way as a human can identify an apple with slightly different features from any apple they have seen before, so too can a trained NN make informed decisions about an unseen data set. This "mental agility" is what separates machine learning from other forms of artificial intelligence.

\begin{figure*}[htb]
   \includegraphics[width=.3\textwidth]{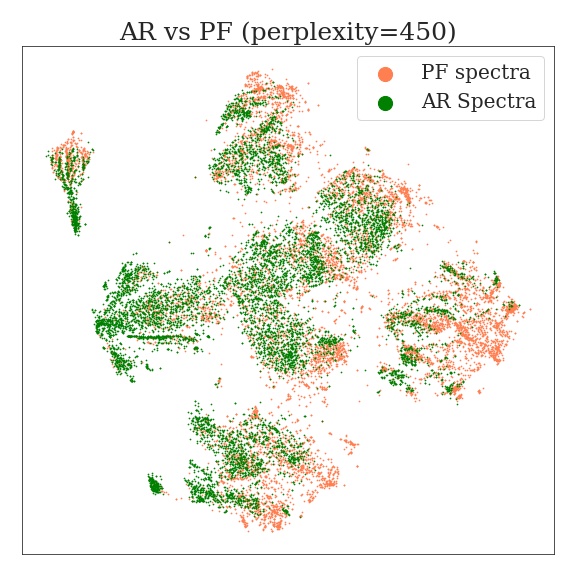}
   \includegraphics[width=.3\textwidth]{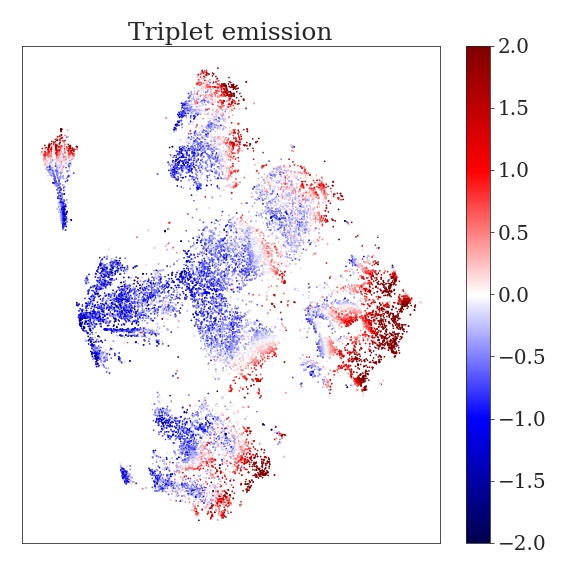}
   \includegraphics[width=.3\textwidth]{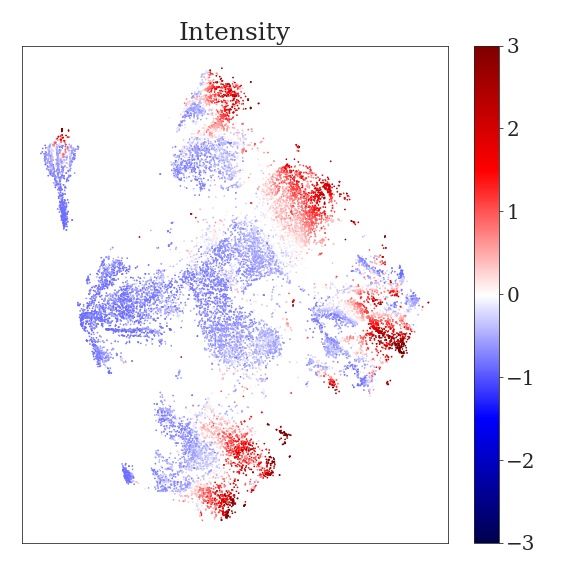}
   \includegraphics[width=.3\textwidth]{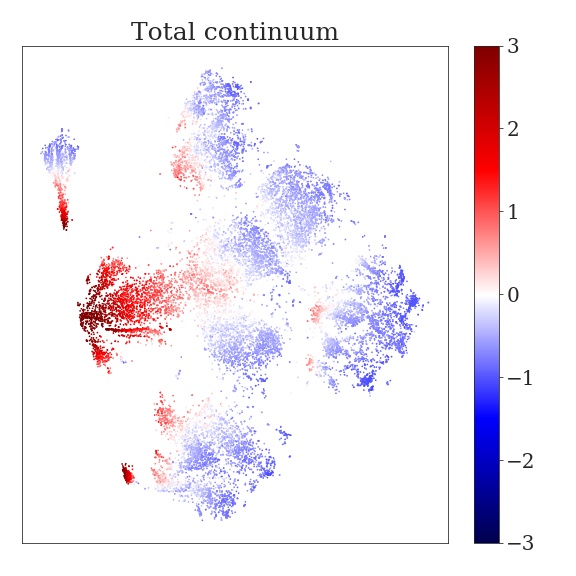}
   \includegraphics[width=.3\textwidth]{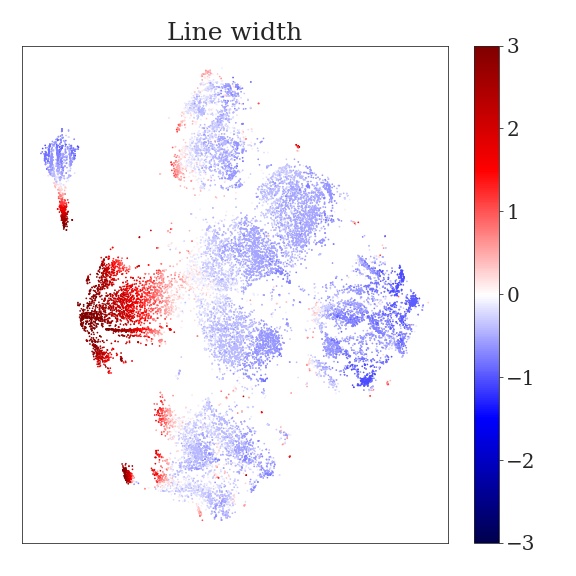}
   \includegraphics[width=.3\textwidth]{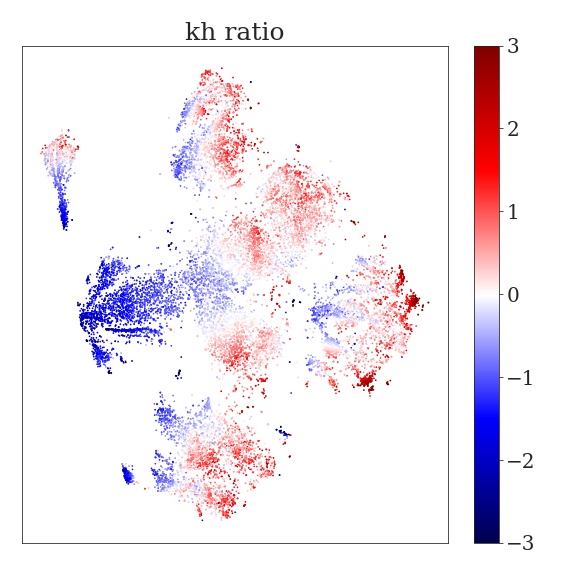}
   \includegraphics[width=.3\textwidth]{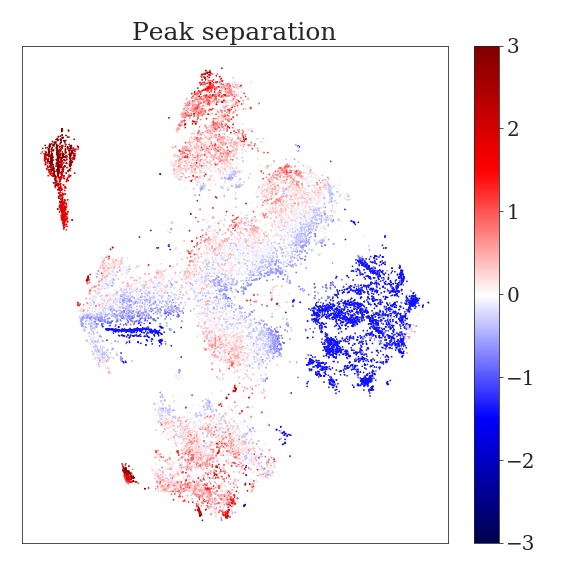}
   \includegraphics[width=.3\textwidth]{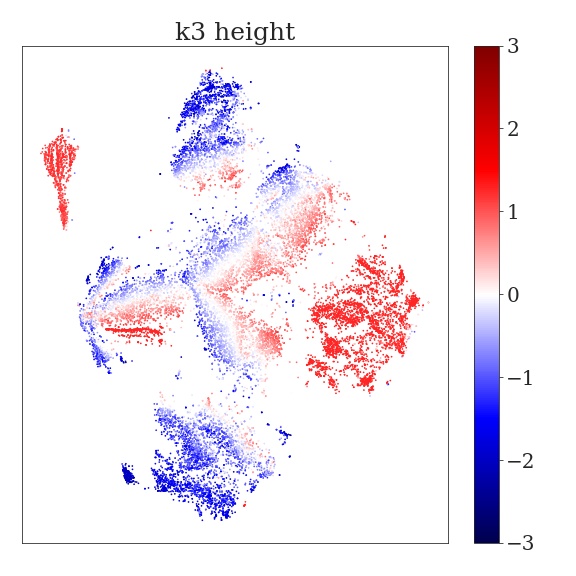}
   \includegraphics[width=.3\textwidth]{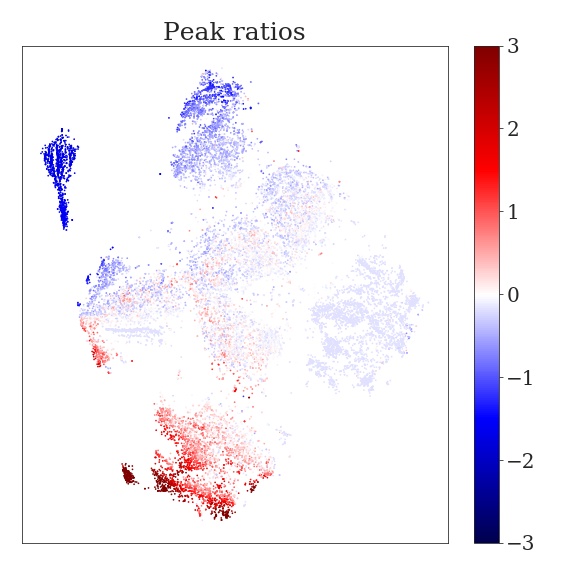}
   \includegraphics[width=.3\textwidth]{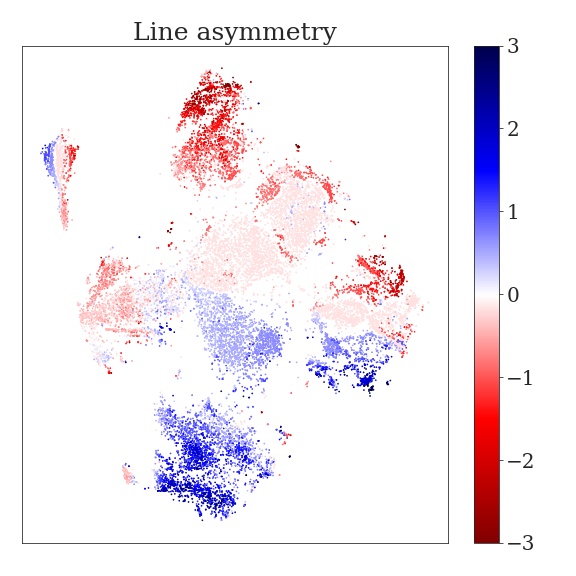}
   \centering
  \includegraphics[width=.3\textwidth]{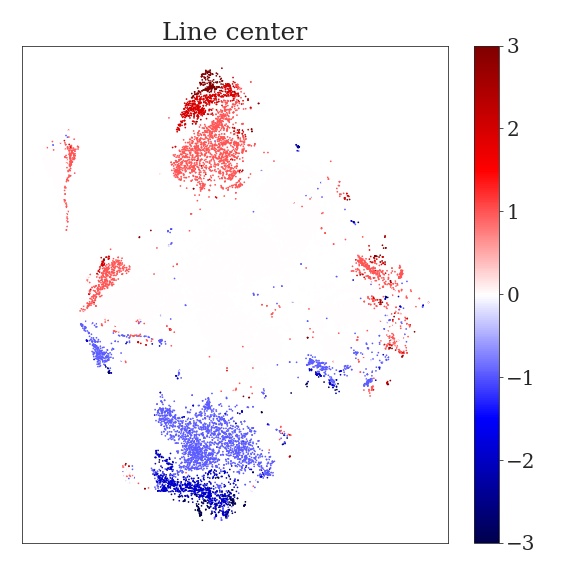}
\caption{Projection plots of the standardized value of each spectrum's 10 features. The upper left panel indicates whether a spectrum belongs to the PF (orange) or AR (green) data set. The feature plots have been ordered in terms of their predictive capacity, with the upper plots being most descriptive of the division between PF and AR spectra, and the lower plots least descriptive. Triplet emission seems to be strongly correlated to the spectrum's class, which is in agreement with Figure \ref{feature_distributions}. In the text, we refer to a feature plot such as triplet intensity, as having a horizontal symmetry, and plots such as line center as having a symmetry along the vertical axis.}
\label{feature_bars}
\end{figure*}

\begin{figure}[htb]
    \includegraphics[width=.5\textwidth]{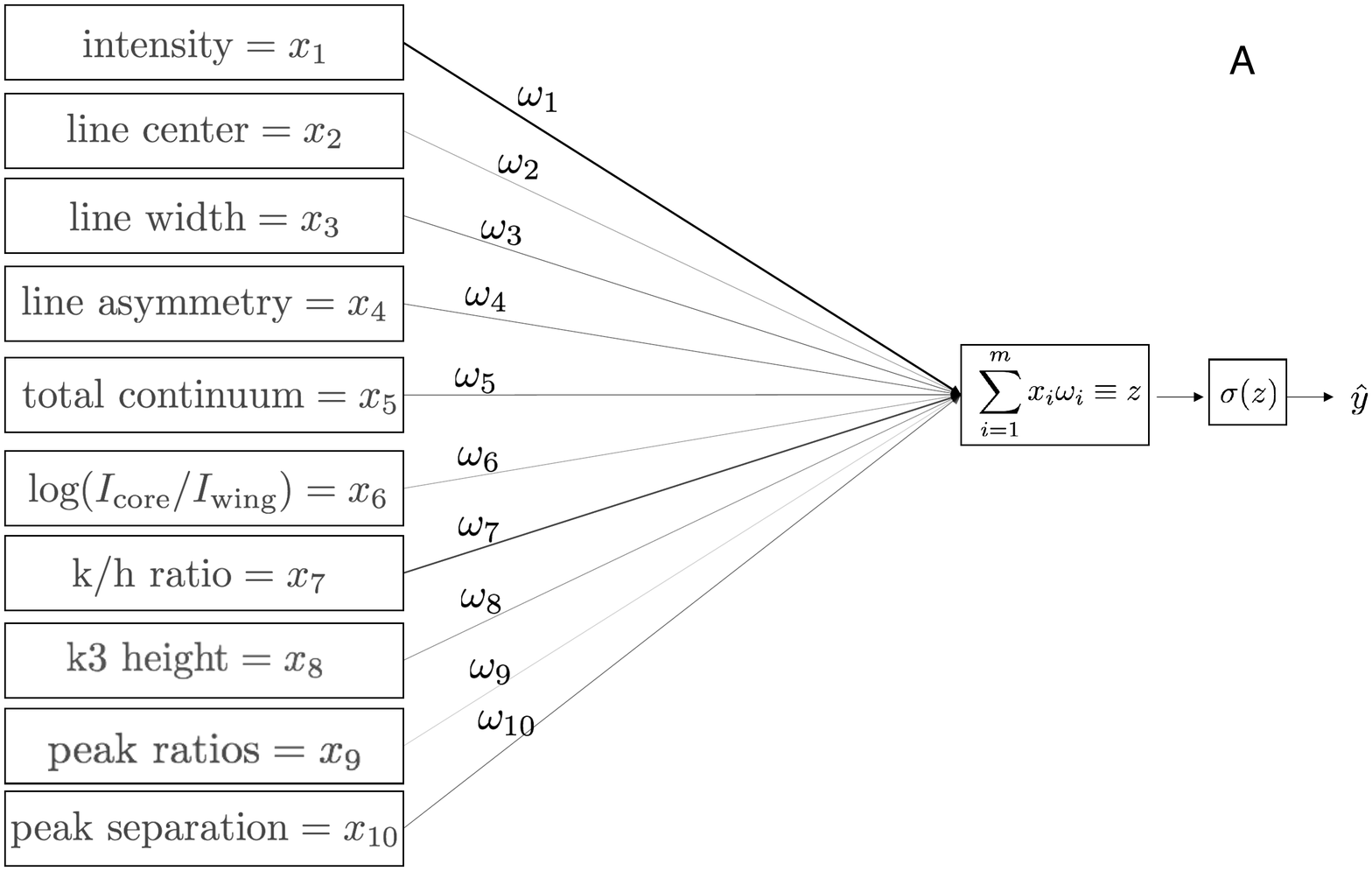} 
    \includegraphics[trim={5cm 0.5cm 0cm 2cm},clip, width=0.3\textwidth]{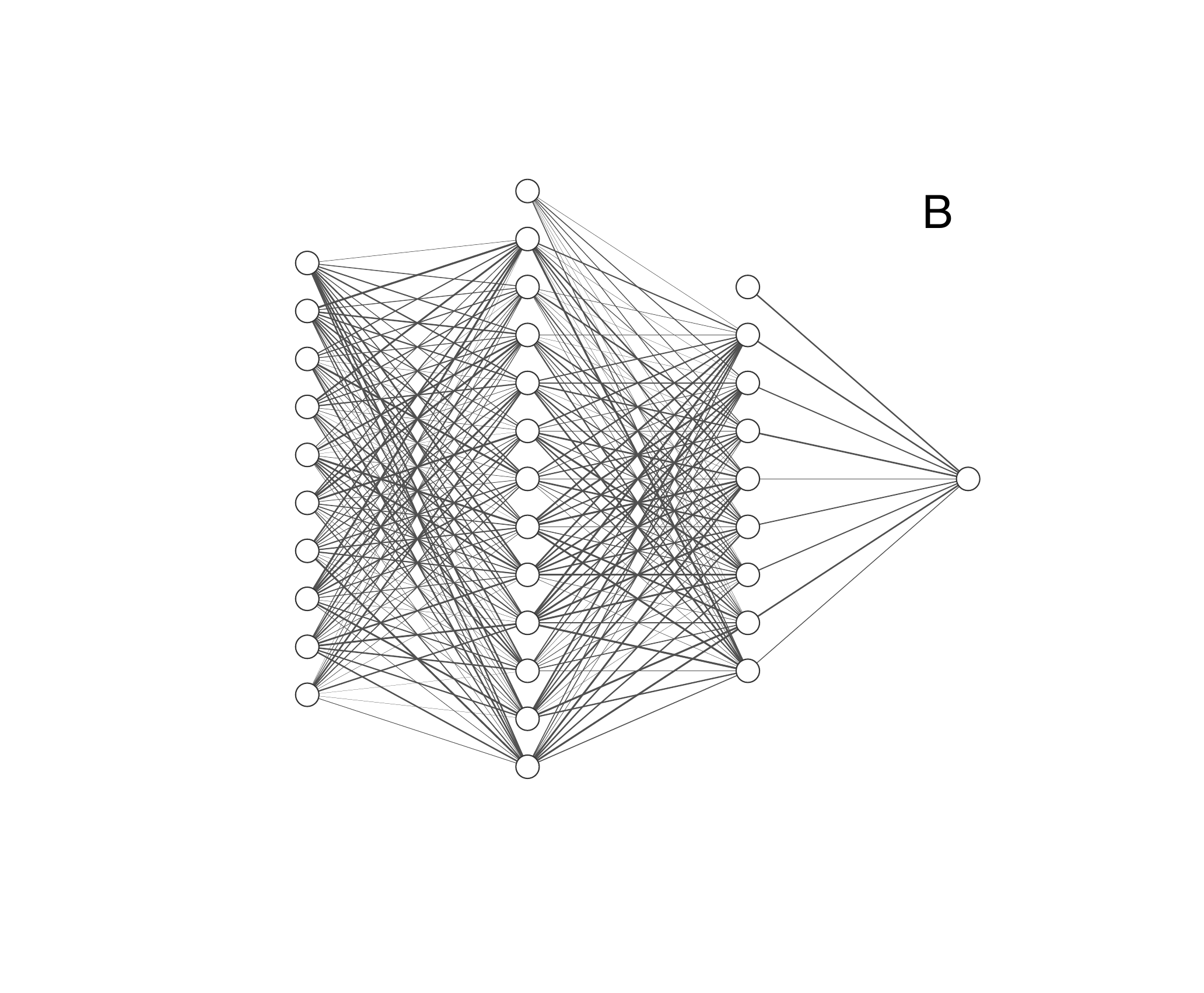} 
    \includegraphics[trim={1cm 1cm 0cm 2cm},clip, width=0.3\textwidth]{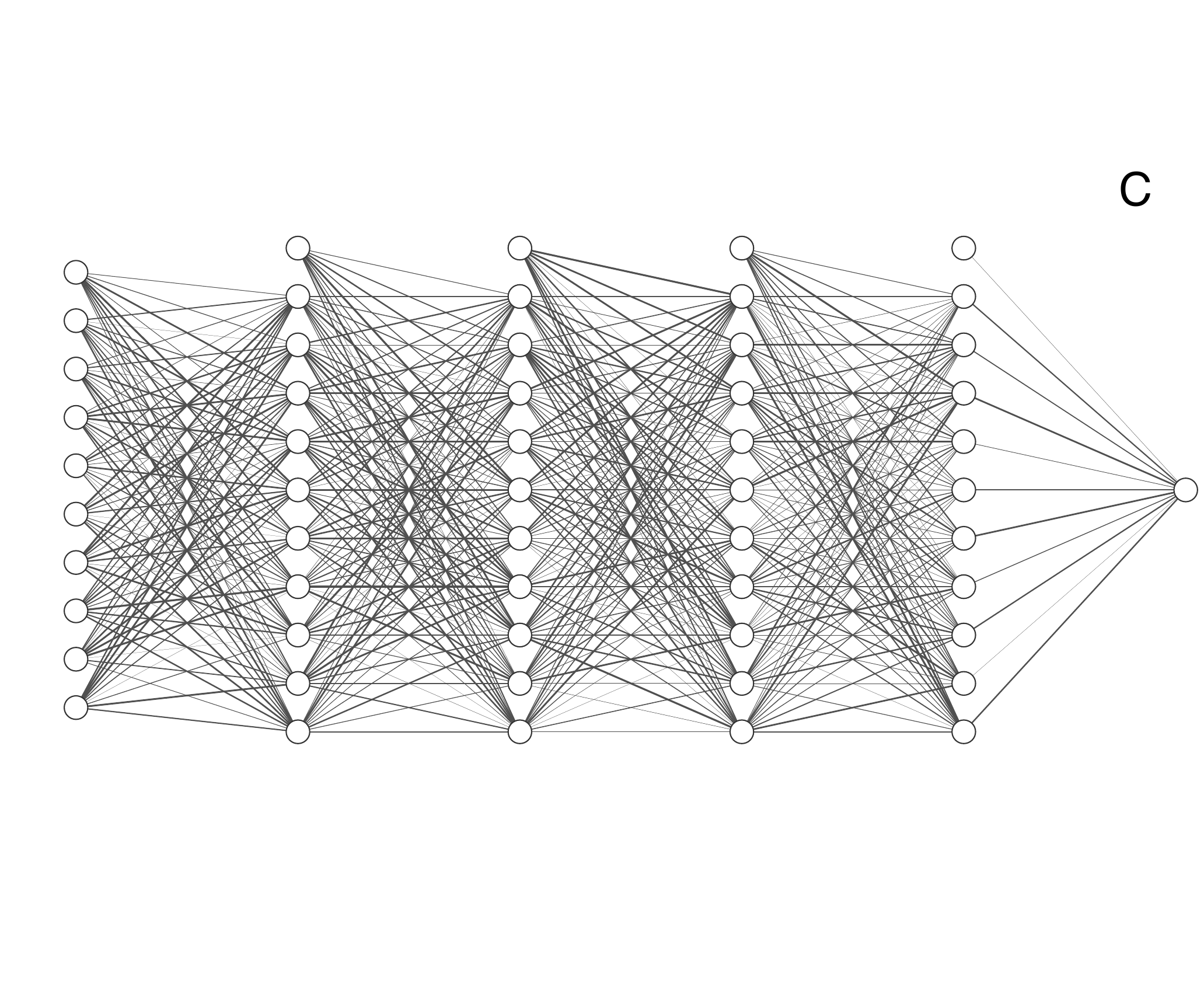}
\caption{Plot of all used fully connected NN architectures. Architecture A has no hidden layers and simply runs a linear combination of the input features through an activation function. Such a network is called a \textit{perceptron}, and is the simplest possible NN. Architectures B and C include multiple hidden layers with additional bias terms (nodes on top). These terms are not connected to the input and introduce an additional set of parameters used to regulate the baseline sensitivity of nodes in each successive layer. All nodes use a rectified linear activation function $f(x) = x^+ = max(0, x)$, except for the last node of each network, which uses a sigmoid function $\sigma(z) = 1/(1-e^{-z})$ to convert the inputs into a probability between 0 and 1: 1 for PF, 0 for other. The thickness of the lines are randomly selected to illustrate weights $w_{ij}$ of different magnitude connecting each node. For instance, imagine model A has been trained and converges to a set of weights whose magnitudes are proportional to the thickness of the lines, in this case, intensity would be more important than peak ratio when identifying PF profiles.}
\label{nn_arch}
\end{figure}

The upper image in Figure \ref{nn_arch} shows a schematic of the simplest type of NN called a \textit{perceptron}. This NN takes the 10 descriptive features associated with a single profile, and forms a linear combination assigning relative importance to each feature through a set of pre-factors called \textit{weights} $w_i$. This linear combination is then passed to an \textit{activation function}, in this case a sigmoid function $\sigma(z) = 1/(1-e^{-z})$, which outputs a $\hat{\text{y}}^i$ value between 0 and 1. The linear combination plus activation function is called a node, and in NN diagrams is commonly represented by a circle. If a profile $x^i$ belonging to the PF data set is fed into the NN, and the NN predicts that the profile comes from one of the other three data sets, i.e., produces an incorrect output close to $0$, then the NN makes corrections to the weights $w_i$ through a process known as \textit{back propagation}, which allows the network to perform gradient descent across some cost function $\mathcal{L}(\hat{\text{y}}^i, \text{y}^i)$ sensitive to the severity of the mismatch between the guessed label $\hat{\text{y}}^i$ and the actual label $\text{y}^i$. Profiles from the training set are fed through the network and the weights are continually adjusted until the model starts understanding the difference between PF/non-PF profiles. If there is no difference between these two classes, then the NN will converge to a model which always outputs a value of 0.5.

The perceptron forms the basis of more complicated NNs in the same way that a biological neuron represents the building block for an organic brain. For more complicated NNs, the input features could connect to multiple nodes, each with their own associated system of weights such as those in the lower two panels of Figure \ref{nn_arch}. NNs that have two or more \textit{hidden layers} (layers not including the input and output layer) are commonly referred to as \textit{Deep Neural Networks}, with each layer learning a more abstract internal representation of the data. The number of layers in a network, number of nodes per layer and particular activation functions define what is called the network's \textit{architecture}. It is difficult to know what architecture best suits a particular problem, and therefore the architecture is chosen based on trial and error. For a more comprehensive review about the general mathematics behind NNs, we direct the reader to the detailed book of \cite{ML_book}.

We tested three NN architectures: architectures A, a simple perceptron with 0-hidden layers, architecture B with 2-hidden layers and architecture C with 4-hidden layers, as seen in Figure \ref{nn_arch}. We also included an SVM with a non-linear kernel. SVMs are linear classifiers used for binary classification problems, which can be made to fit non-linear \textit{decision boundaries} by mapping the data into a higher dimensional space via the use of an appropriate kernel function. An SVM of this type should be comparable with a 2-hidden layer NN, where the first layer projects the data into a more abstract space, while the second layer classifies the data. For a detailed description of an SVM applied to solar flare prediction, see for example \cite{Monica_HMI}. The three models in Figure \ref{nn_arch} are what we call \textit{fully connected} NNs, on account that every node between adjacent layers is connected.

\subsection{Performance metrics}\label{metrics}
Each model is trained on a fraction of the data, and the model's performance is measured on the test set. Several different performance metrics can be obtained by comparing the actual set of labels $\text{y}_\text{test} = (0,1,1,\cdots, 1,0,1)$ to the model's output guesses $\hat{\text{y}} = (.1,.8,.6,\cdots,.7,.2,.9)$. The results of a binary classification problem can be characterized by a \textit{confusion matrix}, where the number of true positives $\mathrm{TP}$ ($\text{y}^i=1~\vline~\hat{\text{y}}^i=1$), false negatives $\mathrm{FN}$ ($\text{y}^i=1~\vline~\hat{\text{y}}^i=0$), true negatives $TN$ ($\text{y}^i=0~\vline~\hat{\text{y}}^i =0$) and false positives $\mathrm{FP}$ ($\text{y}^i=0~\vline~\hat{\text{y}}^i=1$) are used in different combinations to express different aspects of the model's performance. We list all of the metrics used in this study, starting with
\begin{equation}
    \text{precision} = \frac{\mathrm{TP}}{\mathrm{TP} + \mathrm{FP}},
\end{equation}
which is a measure of reliability when the model predicts that a profile comes from the PF data set.
\begin{equation}
	\text{recall} = \frac{\mathrm{TP}}{\mathrm{TP} + \mathrm{FN}},
\end{equation}
on the other hand is a measure of how many PF profiles were correctly labeled. Looking at just one of the above two metrics could be misleading, since it is possible to have a high precision and a low recall at the same time. The harmonic mean of both measures addresses this ambiguity by rolling both metrics into the so called F1 score, given by  
\begin{equation}
	\text{F1} = 2\left(\frac{\text{precision} \times \text{recall}}{\text{precision} + \text{recall}}\right).
\end{equation}
Notice that this formulation gives an interpretation  of performance based on the coupled behavior of both quantities, i.e., favorable performance is assigned to classifiers/models who have both high precision and recall scores.  Another useful metric is the ratio of correct predictions over the total number of predictions, called the accuracy: 
\begin{equation}
	\text{accuracy} = \frac{\mathrm{TP} + TN}{P + N},
\end{equation}
where $P$ and $N$ are the number of observations in the positive and negative class respectively. These measures are susceptible to misinterpretation due to the ambiguity introduced by the relative size of each class, called the class imbalance. For instance, say we have a high class imbalance with only a single PF profile in comparison to 99 non-PF profiles. Labeling every profile as non-PF results in a deceptively positive accuracy of 99\%. To address the problem of class imbalance, \cite{TSS_sugest} suggest the use of a ratio invariant skill score \citep{ratio_invariant}, known as the \textit{true skill statistic} (TSS). This measure uses all of the confusion matrix elements and promotes a standard for the comparison of studies with different flare/no-flare ratios. The TSS is defined as the difference between the recall and false alarm rate:
\begin{equation}
TSS =\frac{\mathrm{\mathrm{TP}}}{\mathrm{\mathrm{TP}}+\mathrm{\mathrm{FN}}}-\frac{\mathrm{\mathrm{FP}}}{\mathrm{\mathrm{FP}}+\mathrm{TN}},
\end{equation}
with a TSS score of 1, 0 and -1 indicating a model with perfect, random and adverse labeling respectively. The problem of class imbalance is not directly applicable to our case, since we ensured that the number of spectral profiles in the PF/non-PF classes were held equal at all times, however, for completeness we have included as many measures as possible.\\
 
\begin{figure*}[tbh] 
\centering
\includegraphics[width=1\textwidth]{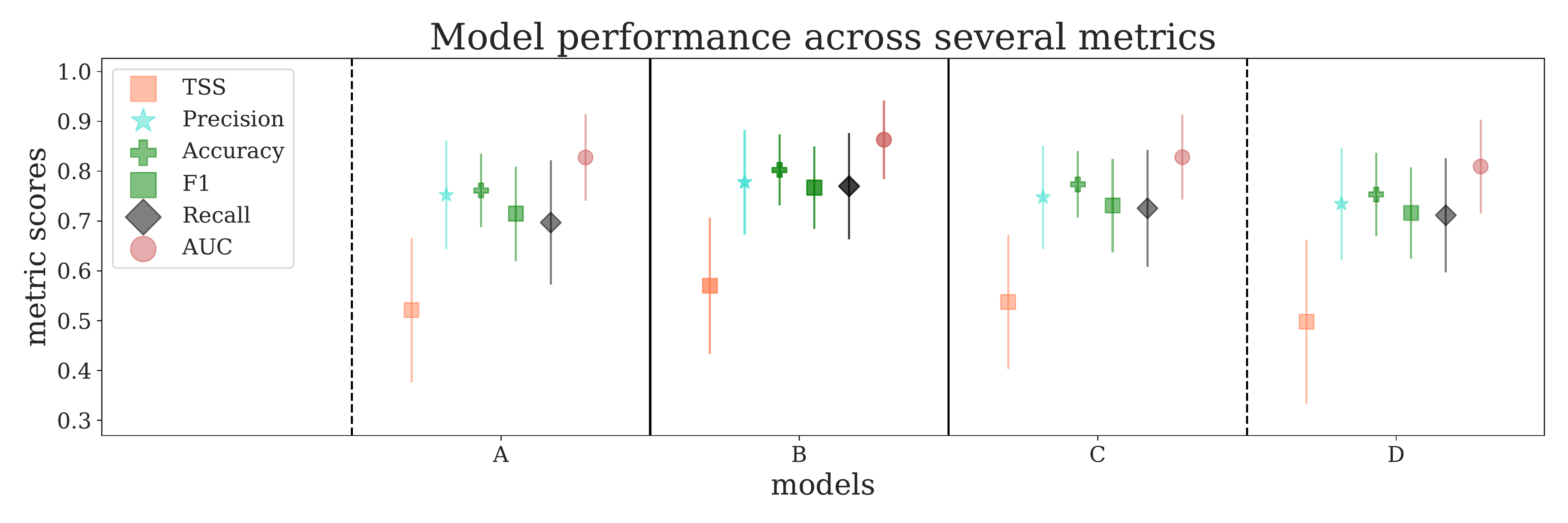} 
\caption{Performance of models A-C (shown in Fig. 11), and D (an SVM) based on 6 binary classification metrics. Each metric has a favorable maximum score of 1, and a random guess value of .5, except for the TSS which has a floor of -1. The 2-hidden layer NN corresponding to model B has a superior performance across all metrics. The scores were derived from a binary classification problem AR vs PF over an observationally split data set. The scores where averaged over 100 iterations of randomly split training and testing data sets, with error bars representing one standard deviation of each metrics score.\label{metric}}
\end{figure*}

The performance of each model over the binary classification problem PF/AR, can be seen in Figure \ref{metric}, with each column corresponding to the performance of a specific model on an \textit{observationally split} data set. The term "observationally split" means that each IRIS observation has either been isolated to the training or test set, with none of that observation's spectra being shared between the two data sets. If the training and test sets were divided on a spectral basis only, without regard for the division of observations, the models could gain leverage on the shared data, and learn spectral peculiarities unique to each particular observation. Splitting our data sets observationally ensures that our results are not artificial and remain as close to reality as possible. Figure \ref{metric} shows that architecture B, with two hidden layers, has the strongest performance across all metrics. The splitting of data into a training and test set injects a level of uncertainty into the results, with different splittings resulting in slightly different model performances. In machine learning, an estimation of this uncertainty can be captured by performing multiple runs with the same model, whilst continually swapping out the observations that appear in the test and training sets. Therefore, to construct the error bars in Figure \ref{metric}, we performed the computationally expensive exercise of 1) Randomly splitting the AR and PF observations seen in Table \ref{qsssar} and \ref{pf}, such that each random splitting resulted in the delegation of 4 AR and 4 PF observations in the test set, whilst the spectra from the remaining observations formed the training set. 2) Training each model architecture over 15 epochs (the number of times the network is allowed to see the entire data set to update its weights) before extracting the model at an epoch which scored the highest accuracy on the test set. 3) Calculate the model's performance in terms of all the above mentioned metrics over the data from the current unseen test set. This three step procedure was iterated 100 times for each model, resulting in a distribution of each model's metric scores. The mean of each metric score was then plotted in Figure \ref{metric}, and the standard deviations were used as error bars.

\begin{figure}[htb] 
\centering
\includegraphics[width=.49\textwidth]{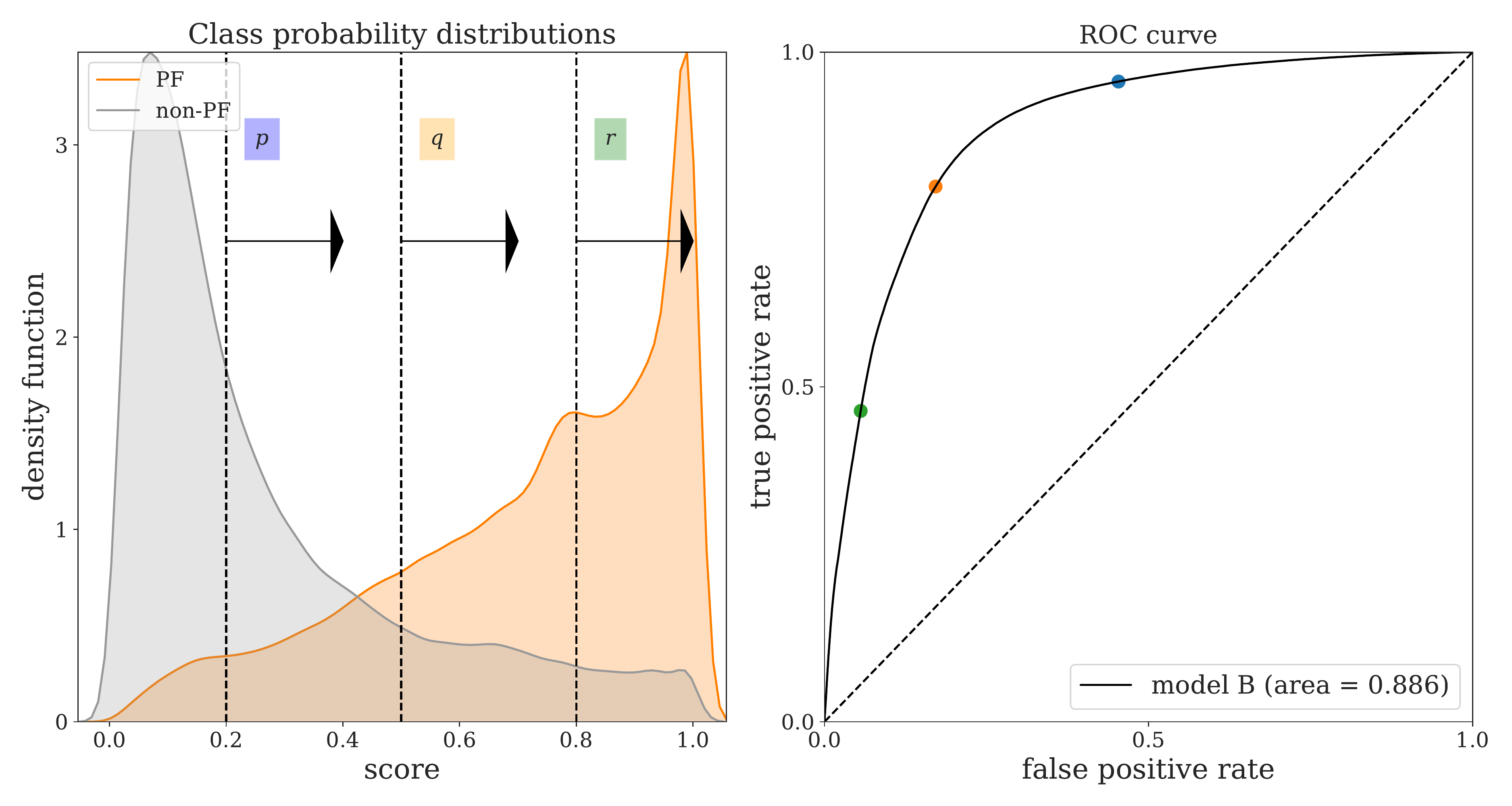}
\caption{The left panel shows the probability distributions assigned by model B to the AR/PF classification problem with respect to an unseen observationally split test set. The vertical dashed lines labeled p, q and r, indicate the effect a shift of the threshold value has on the true positive and false negative rates as seen on the ROC curve to the right. Selecting the threshold value at vertical line p, results in the majority of PF spectra being correctly labeled along with a large amount of AR spectra being incorrectly labeled, this in turn resulting in a high true and false positive rate. The overlap between the two distributions can be used as an inverse measure for the separability of the AR and PF data sets, and is characterized by the area under the ROC curve.}
\label{roc_obs_split}
\end{figure}

Each of the metrics mentioned thus far are sensitive to the so called \textit{threshold value} $\delta$, which defines the "watershed" between a PF/AR decision. In general, the threshold is set to $\delta=.5$, so that output values $\hat{y} <.5$ are classified as non-PF, while values $\hat{y}\geq .5$ are classified as PF profiles. Choosing a threshold value turns a model into a binary classifier. Adjusting the threshold has a direct effect on the confusion matrix elements, which in turn cascade into all of the performance metrics. For this reason, an additional metric which integrates over the threshold value is preferable. This is commonly achieved by monitoring the relationship between the true and false positive rates given by
\begin{equation}
\begin{split}
\text{TPR}(\delta) &= \int^\infty_\delta f_1(y)~dy, \\
\text{FPR}(\delta) &= \int^\infty_\delta f_0(y)~dy,
\end{split}
\end{equation}
where $f_1$ and $f_0$ are the probability distributions assigned by the model to the positive and negative classes respectively. Both of these distributions can be seen in the left panel of Figure \ref{roc_obs_split}, with $f_0$ (grey) corresponding to the probabilities associated with AR spectra, and $f_1$ (orange) to the PF spectra of a never seen before observationally split test set. As the threshold value changes, so too does the true positive and false negative rate, giving rise to the so called \textit{receiver operating characteristic curve}, or (ROC) curve seen in the right hand panel of Figure \ref{roc_obs_split}. Each point on the ROC curve represents a trade off between a high number of false positives and a high number of false negatives, with the dashed meridian line representing a random guess. A perfect score is represented by a point in the upper left hand corner of the parameter space, and the area under the curve
\begin{equation}
AUC = \int^1_{\delta=0} \text{TPR}\left(\text{FPR}^{-1}(\delta)\right)~d\delta,
\end{equation}
is commonly used as a threshold integrated meta statistic to judge the performance of a model on a binary classification problem. In this way, the AUC can be used to quantify the overlap between the two probability distributions, with larger areas corresponding to less overlap and therefore stronger model performance. Because we integrate out the threshold, the AUC score can be interpreted as the probability that the model will rank a randomly chosen PF spectrum higher than a randomly chosen non-PF spectrum. Therefore, the AUC is a metric that is both scale and threshold invariant. For this reason, the AUC scores have been computed and included in Figure \ref{metric}.

\section{Space and time dependent performance}\label{Space and time dependent performance}
\begin{figure}[htb]
\centering
   \includegraphics[trim={0cm 0cm 0cm 0cm},clip, width=.236\textwidth]{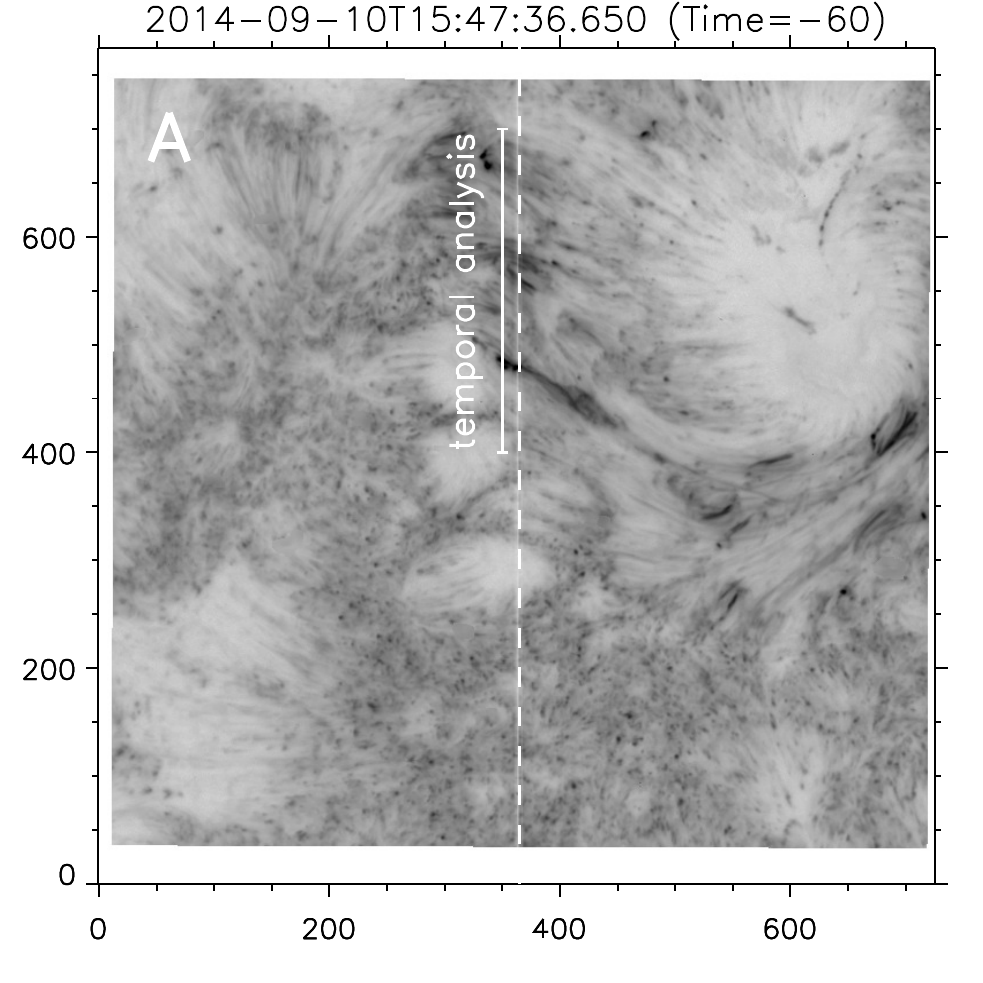}
    \includegraphics[trim={0cm 0cm 0cm 0cm},clip, width=.236\textwidth]{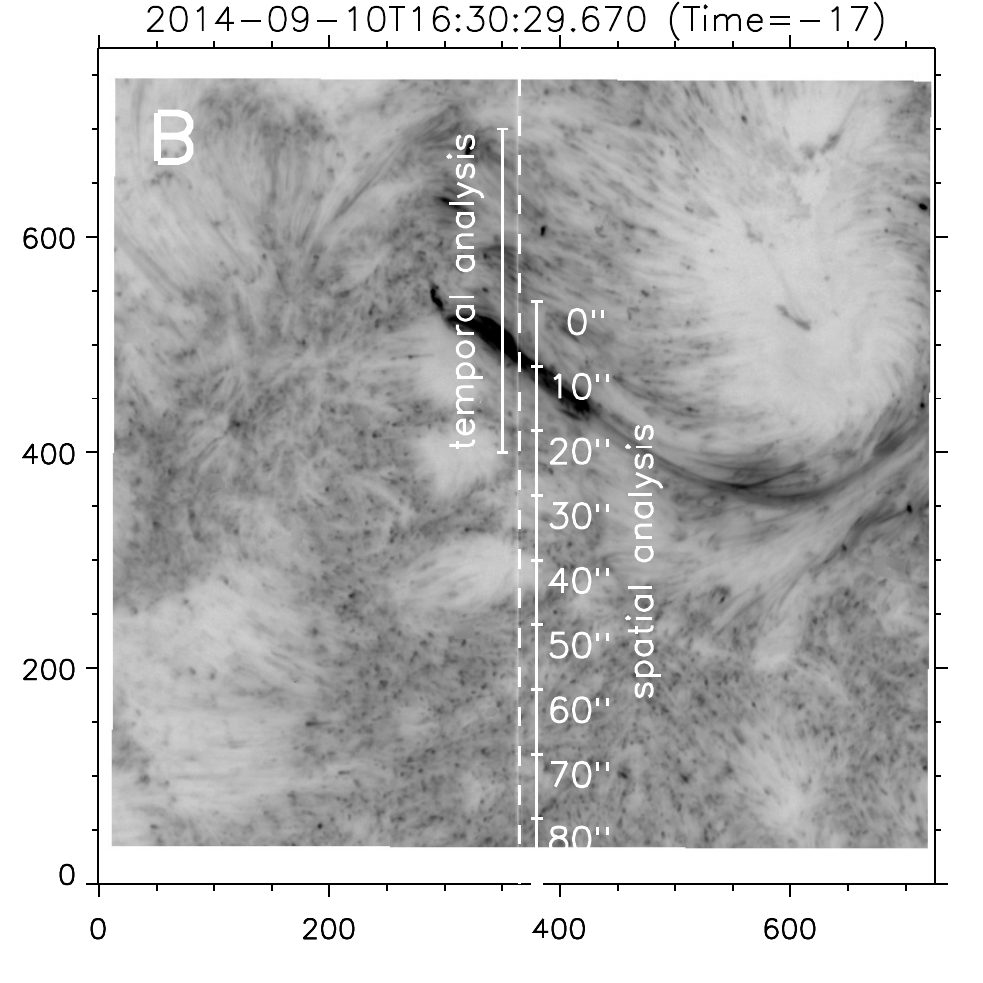}
    \includegraphics[trim={0cm 0cm 0cm 0cm},clip, width=.236\textwidth]{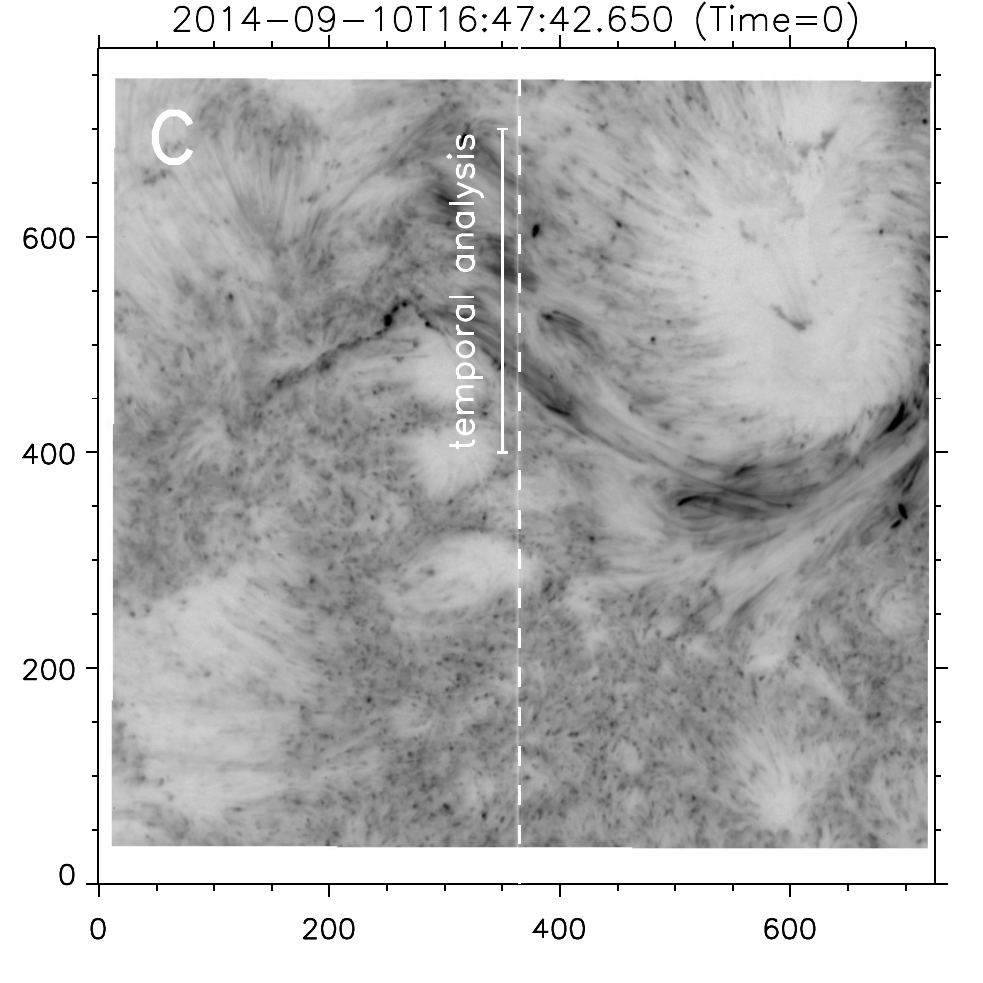}
    \includegraphics[trim={0cm 0cm 0cm 0cm},clip, width=.236\textwidth]{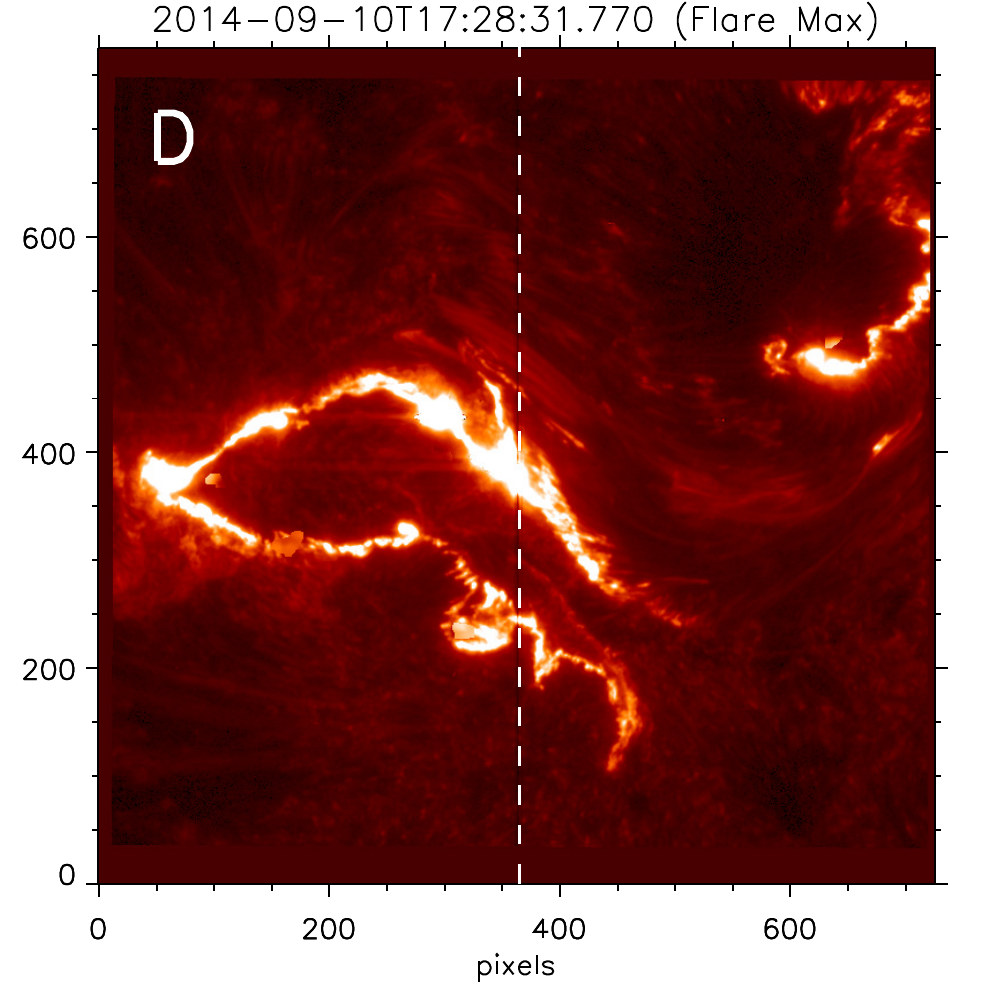}
\caption{SJIs for the temporal and spatial experiment. The SJI in panel A is taken 60 minutes before flare onset. Bright points occur over the upper region of the slit which may account for the increased metric scores. Panel B shows a SJI 17 minutes before flare onset, where the spatial experiment was performed. Panel C shows the state of the active region just before flare onset, and panel D is a SJI taken at flare maximum.}
\label{iris_ims}
\end{figure}

We have trained a model that can distinguish PF from AR spectra, with the majority of AR spectra scoring less than their PF counterparts as shown in Figure \ref{roc_obs_split}. However, all PF spectra up until this point have been sourced from 25 minute observation windows just before flare onset. Arguably, spectra produced closer to flare onset should be easier to identify by the model as PF than those produced over the same active region earlier in time. It is also reasonable to assume that the model will assign higher probabilities to spectra coming from solar regions that display more activity within either the 1400 or 2796 SJI passbands. To test these two hypotheses, we analyze a single PF observation leading up to a two-ribbon X1.6-class flare observed by IRIS on September 10, 2014 in a sit-and-stare observational mode. A snippet of the evolution of each feature for this observation has already been shown in Figure \ref{feature_evolution}. Since we are only looking at a PF observation, all of the spectra are by definition labeled as PF. Consequently, we only have a single probability distribution $f_1$, pertaining to the positive class. The lack of a negative class imposes a symmetry on the confusion matrix elements, resulting in a degeneracy of F1, recall, accuracy and precision metrics. If PF spectra become easier to identify, either by sampling profiles closer to flare onset or from visibly more active regions, then the distribution $f_1$ will shift successively closer to 1, resulting in larger metric scores. SJIs of the PF observation in the 1400 passband can be seen in Figure \ref{iris_ims}.

\begin{figure*}[tbh]
\setlength{\tabcolsep}{0.5pt}
\begin{tabular}{@{}c@{}}
    \includegraphics[trim={0cm 0cm 0cm 0cm},clip, width=1\textwidth]{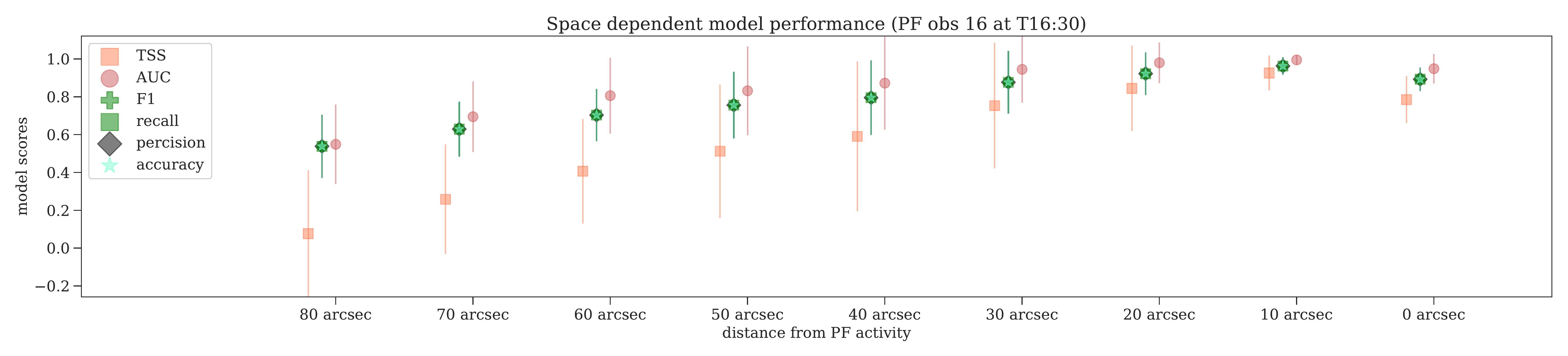} \\
    \includegraphics[trim={0cm 0cm 0cm 0cm},clip, width=1\textwidth]{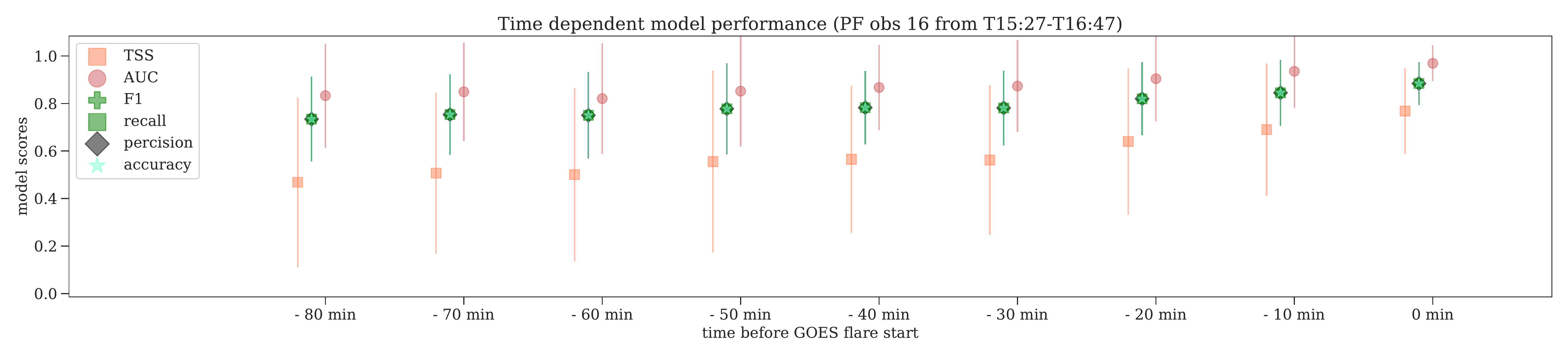} \\
    \includegraphics[trim={0cm 0cm 0cm 0cm},clip, width=1\textwidth]{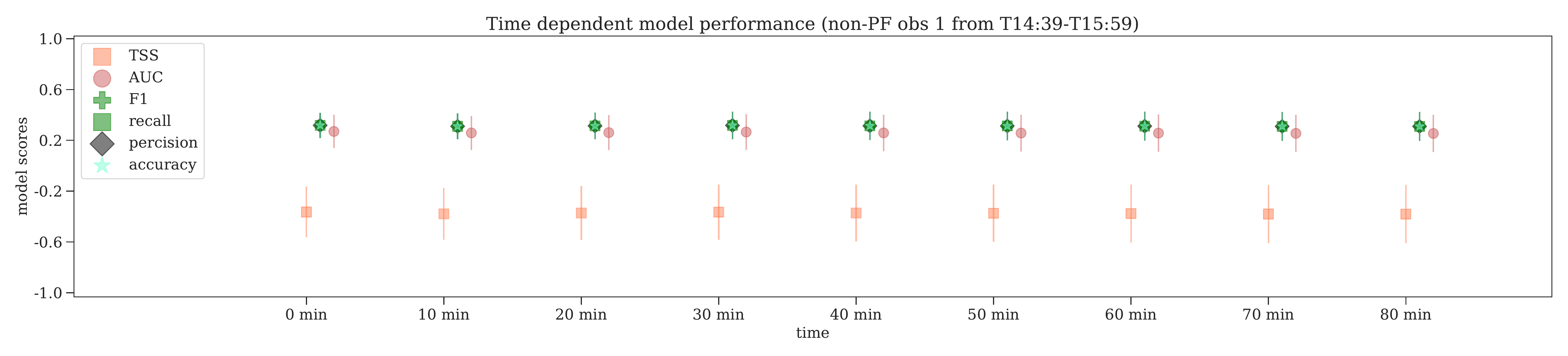} 
\end{tabular}
\caption{Upper panel: Model performance based on distance away from activity in the 1400 passband of a PF active region. The spectra become increasingly PF like, and consequently more of the spectra are correctly labeled as PF for smaller distances away from the brightenings. Note that at sufficient distances away from the PF active region, the model cannot guess whether the spectra come from a PF or AR dataset. Middle panel: Model performance based on time from flare onset. Spectra over a flaring active region become increasingly PF like the closer one is to flare onset. Lower panel: Model performance over a non-flaring active region. The spectra are correctly labeled as non-PF for any time instance. Because the tests are performed on single data sets without their negative class counterparts, many of the metrics express the same measure. The error bars were calculated using an ensemble of model B architectures trained on different splittings of the data sets. The above three observations were never included in the training process.}
\label{space_time_metrics}
\end{figure*}

\begin{figure*}[tb]
    \includegraphics[trim={0cm 0cm 0cm 0cm},clip, width=.33\textwidth]{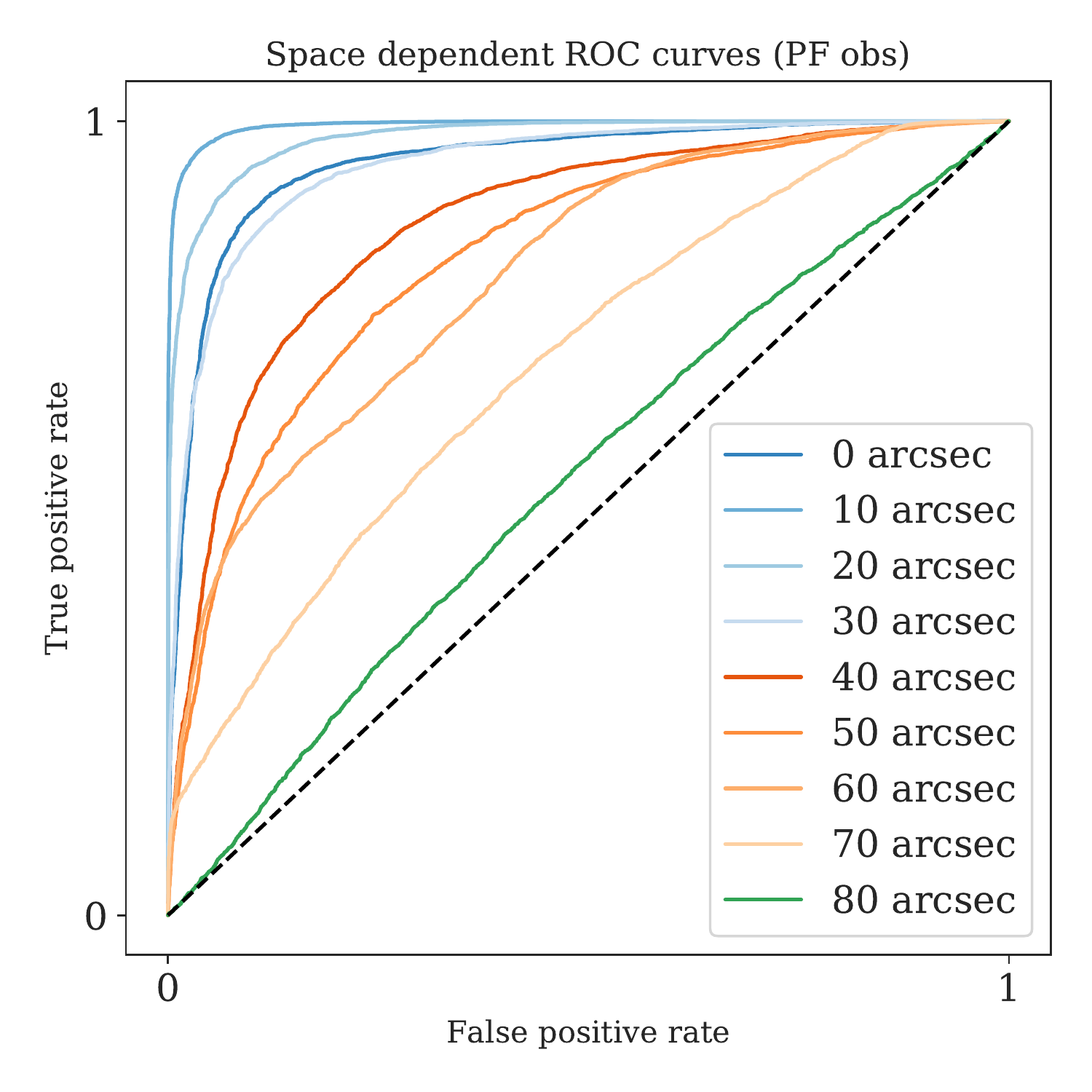}
    \includegraphics[trim={0cm 0cm 0cm 0cm},clip, width=.33\textwidth]{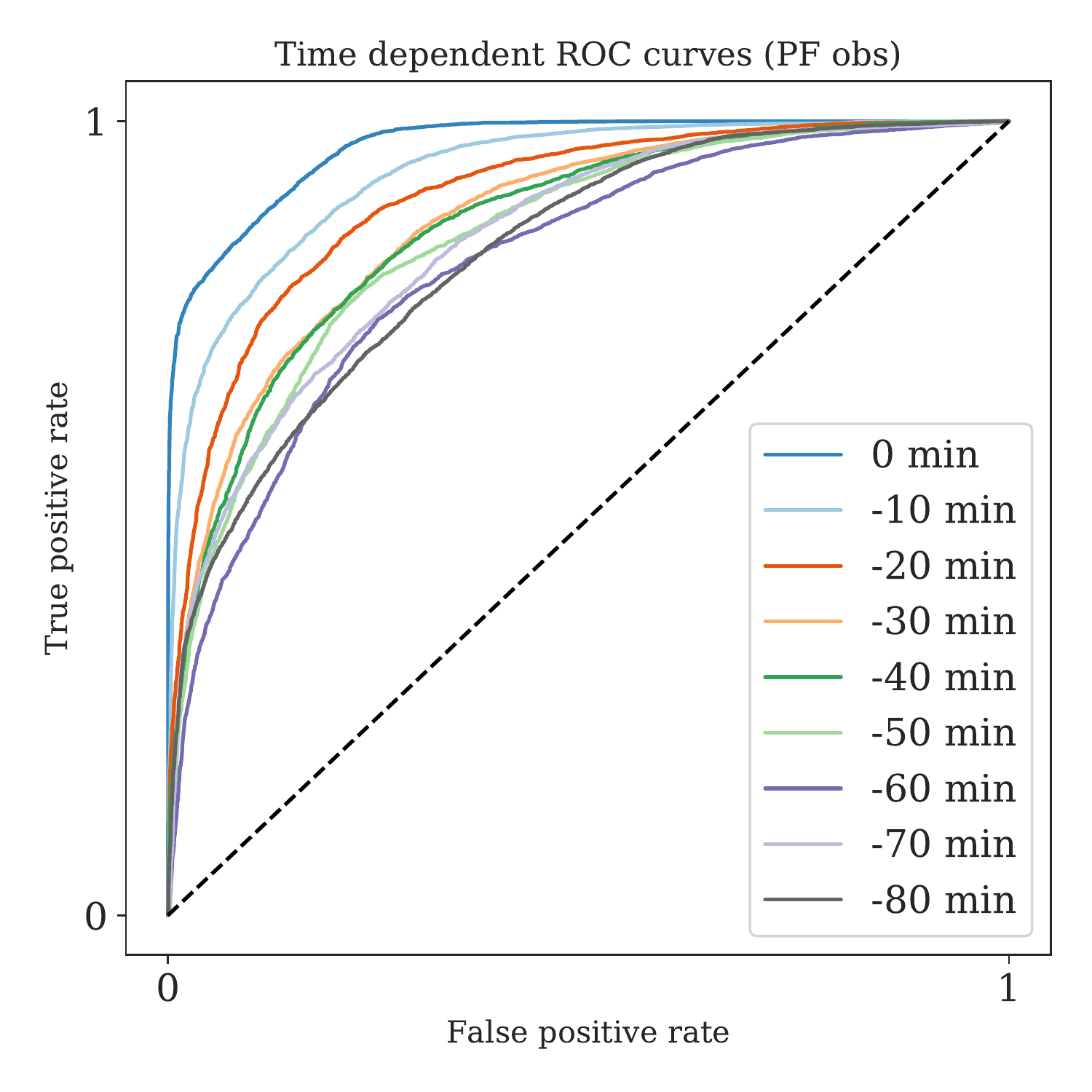}
    \includegraphics[trim={0cm 0cm 0cm 0cm},clip, width=.33\textwidth]{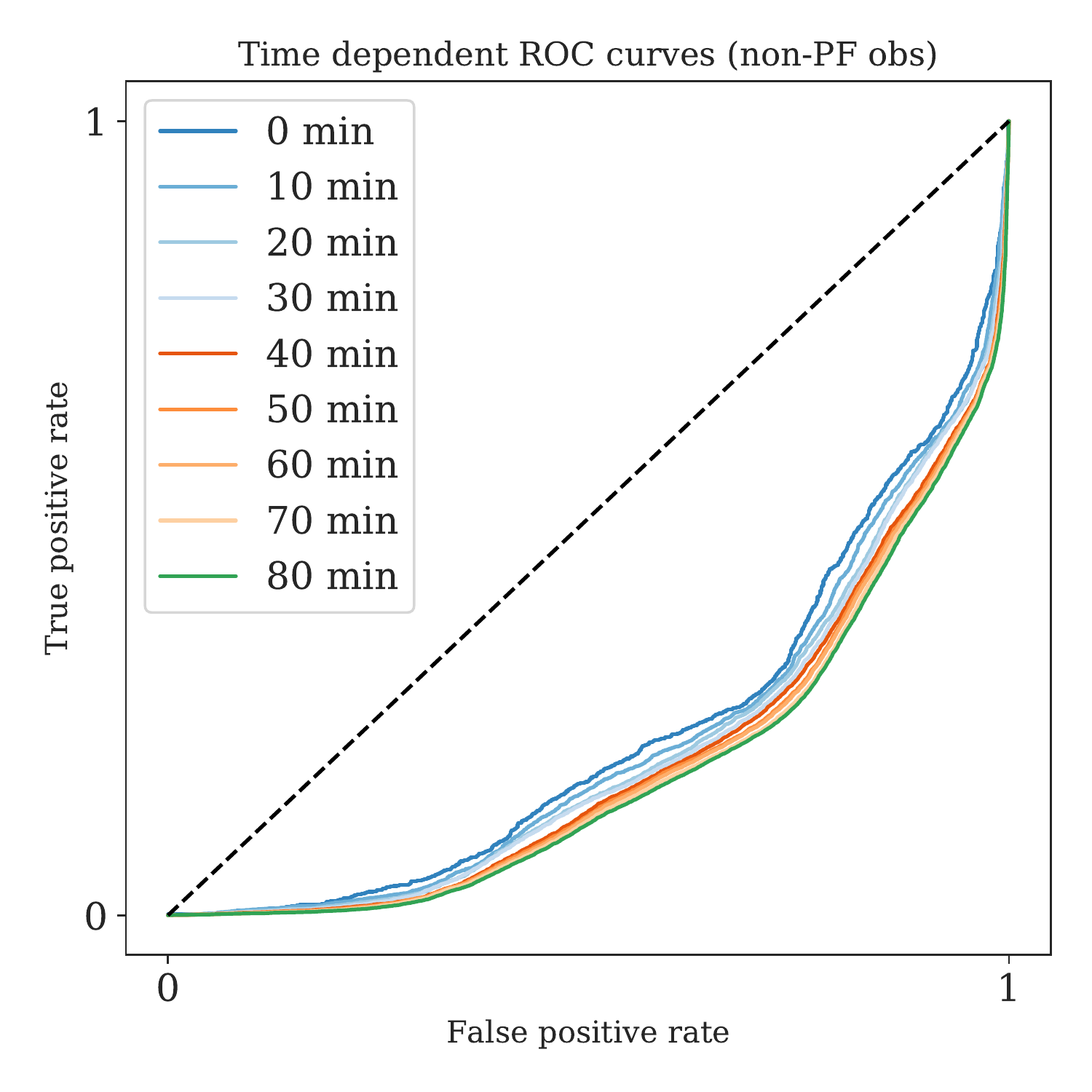} 
\caption{ROC curves corresponding to the results in Figure \ref{space_time_metrics}. Curves with larger areas can be interpreted as the model's level of certainty when making a PF/non-PF judgment, while the orientation about the dashed line (above or below) indicates the models decision (PF or AR) respectively. The first two panels indicate increased confidence in PF labeling when the PF active region is either approached on the solar disk or allowed to encroach on the flare onset time, while the last panel indicates that the model can identify AR spectra as non-PF.}
\label{roc_final}
\end{figure*}

For the spatial analysis, we sourced spectra from eight 10 arcsec bins at a fixed time (17 minutes) before flare onset, as seen in panel B of Figure \ref{iris_ims}. The metric scores and ROC curves associated with a movement from 80'' towards a brightening in the SJI can be seen in the first panels of \Cref{space_time_metrics,roc_final}. For the temporal analysis, we sourced spectra from a window from pixels 400-700 along the slit over 10 minute intervals, starting from 80 minutes before flare onset.  This range of pixels was selected on account of the heightened activity observed in the 1400 SJI passband in this region. The temporal evolution of the metric scores and ROC curves associated with the movement forward in time towards flare onset can be seen in the second panels of \Cref{space_time_metrics,roc_final}.

It is evident that the $f_1$ probability distributions shift towards higher values for both the spatial and temporal experiment. The latter case implies that our model can be used as a rudimentary flare prediction technique. Unlike previous flare prediction algorithms, this approach affords us the opportunity to continually monitor the probability of a flare with every spectral readout, providing the first (to our knowledge) real-time approach to flare prediction.

We also performed a negative test on AR 1 in Table \ref{qsssar}. Despite showing large intensity enhancements in the 1400 passband, this observation never lead to a flare. The models performance can be seen in the last panels of \Cref{space_time_metrics,roc_final}. In this case, we expect the $f_0$ probability distribution associated with the negative AR class to be shifted closer to 0. The metric scores and ROC curves indicate that this is indeed the case, implying that the model has managed to correctly label active region spectra for the entire observation.

\section{Discussion}
We set out to discover whether \ion{Mg}{2} spectra generated in QS, SS, AR and PF observations are distinguishable from one another, with a special emphasis on the last two regions as they may pertain to flare prediction. We now examine the supporting evidence in favor of separability. 

\subsection{Feature distributions}
The first clue to support the notion of separability can be found in Figure \ref{feature_distributions}. Many of the feature distributions show distinct differences between the four data sets. The QS profiles for instance have on average, higher continuum emission, lower triplet emission, deeper central reversals and appear to be formed under comparatively optically thicker conditions as suggested by their reduced k/h ratios. On the other hand, the distinguishing features of PF profiles appear to be enhanced intensities, triplet emission and decreased continuum emission. We caution the reader when making literal inferences from these distributions, as they are subject to both the performance of our feature finding algorithm, as well as the particular definition of each feature, which in the case of triplet emission can be quite obscure.

\subsection{Low-dimensional embeddings}
low-dimensional representations of the data provide an alternative but complementary approach to viewing the possible separability of the four data sets, with both PCA and t-SNE supporting the premise that PF profiles can be distinguished from QS, SS and AR profiles. The t-SNE representations in Figure \ref{t-SNE plots} appear to display more dissimilarity than their PCA counterparts, while at the same time maintaining richer structures that may be indicative of the original data set. There is a clear distinction between QS and PF profiles for perplexities in excess of 2. The distinction between SS, AR and PF profiles is slightly more strained, with a greater overlap between profiles from different data sets, however, profiles from different regions tend on average to occupy distinct subspace, especially for higher perplexities.

Increasing the perplexity from left to right results in an evolution from noisy circular representations with the two relevant classes being largely indistinguishably distributed (first column), to more ordered and complex structures. One can understand this evolution as follows: The perplexity defines the volume of space that a particular data point is aware of. The larger the perplexity, the more neighbors are included in each point's reference frame. t-SNE then embeds the data into a low-dimensional space, such that each point's frame of reference is minimally augmented. In this way, the data points can be distorted on a global scale, while still maintaining the same local structure. The first column in Figure \ref{t-SNE plots} has the lowest perplexity, meaning each point is surrounded by a small volume containing only its nearest neighbors. Consequently, the global structures such as the data's border are of no significance to the map, and the representation only captures local clusters. In contrast, large perplexities have to preserve large scale structures, leading to more elaborate borders.\\

Beyond their immediate aesthetic appeal, low-dimensional embeddings often serve no purpose other than to display the separability of classes or speed up the training of certain models by removing superfluous features. Projecting feature values onto the final embeddings, like we have done in Figure \ref{feature_bars}, allow us to gain traction back into the world of physics. We find it best to analyze Figure \ref{feature_bars} in terms of symmetries. For instance, triplet emission has a rough symmetry about the horizontal axis, corresponding to the maximum direction of change of the triplet emission variable. In contrast, the line center divides the plot in such a way that it has an approximate symmetry about the vertical axis. Since the reference image in the upper left plot indicates a symmetry about the horizontal axis, any feature with horizontal symmetry is more useful for class prediction, because the direction of change of said feature overlaps maximally with how the PF and AR profiles are distributed. Following this line of reasoning, in order of predictive capacity we have: triplet emission, followed by intensity, total continuum, line width, k/h ratio, peak separation, k3-height, peak ratios, line asymmetry, and finally, line center position. Conversely, those features displaying symmetry about the vertical axis are leased valuable as class predictors, since profiles from either the PF or AR class can be found anywhere along their range. This is the case for all velocity features, which clearly divide the upper and lower hemispheres of the plot into down and upflows. For instance, assuming that a mean of 0 (grey) corresponds to an unshifted spectral profile, the feature associated with line centre, (bottom right feature plot), divides the upper and lower hemispheres of the maps into down and upflows respectively. A similar division is made in the line asymmetry map, with red negative values associated with downflows and blue positive values with upflows. The peak ratios divide the map in a similar way, but indicate a reversal of velocities with the lower region being associated with downflows and the upper region with upflows. We are not sure how to resolve this discrepancy, and are hesitant to dismiss it as an oddity arising from our particular methods of measurement. What we can conclude however, is that on average, velocity features seem indistinguishably distributed between PF and AR profiles. All these observations are consistent pictorial realization of the F-scores displayed in Figure \ref{Fscores}.

The t-SNE feature projection plots allow us to examine the coupled behavior of features, something that cannot be directly obtained from the univariate Fisher scores. Features corresponding to opposite symmetry groups represent loosely uncoupled variables. For instance, a profile can be highly symmetric or antisymmetric regardless of how much triplet emission it has. The uncoupled sets of variables are \textbf{horizontal}:(triplet emission, intensity, total continuum, line width, k/h ratio), \textbf{middle}:(peak separation, k3-height), \textbf{vertical}:(line center, line asymmetry, peak ratios). Each feature has been ordered such that the leading features of each group have the highest degree of freedom. In contrast, features which fall into the same symmetry groups are coupled to some degree. An interpretation of the coupling from the horizontal group is as follows: More intense profiles are often accompanied by triplet emission and decreased continuum emission, and have a tendency to form under optically thinner conditions (higher k/h ratios), with a weak propensity towards closer possibly single peaked spectra. Additionally, peak separation and k3-height are strongly anti-correlated, with deeper reversals being associated with broader profiles.

The strong coupling between total continuum and line width should be dismissed as artificial, since it is clear that some continuum is included within the line windows of Figure \ref{profile}. Therefore, an increase in total continuum results in an increase in line width. Once again, these interpretations are based on the performance of our feature finding algorithm, and the statements we have made are statistical in nature, and only capture general trends.

\subsection{Model performance}
From figure \ref{metric}, it is clear that model B is superior on all fronts when it comes to distinguishing PF/AR profiles, with the 4-hidden layer NN performing below expectations. This may be due to overfitting, since the network has many more free parameters which could be used to form extremely complex \textit{decision boundaries} that do not generalize well. However, no such overfitting was apparent from the \textit{learning curves} and subsequent trials involving the insertion of \textit{dropout} layers returned little to no improvement. We can conclude that the division between PF/AR profiles is relatively simple, requiring only a light weight NN to produce satisfying results. The success of model B may also indicate the relevance of our choice of basis, since additional feature learning layers do not seem to be required. We expect that the intrinsic number of dimensions necessary to represent the data is no more than 10. Our basis was selected on the grounds that each feature can be used as a diagnostic for some observable. If a feature cannot be linked unambiguously to some statement about the solar atmosphere, then it is by definition random and non-descriptive.

The high metric scores warrants the conclusion that there is a great deal of continuity between IRIS observations of the same type. That is, different IRIS PF observations produce similar looking spectra.\\

It appears that there is a persistent overlap in Figure \ref{roc_obs_split} between the classification distributions of PF/AR scores. We postulate that this overlap has two distinct sources. The overlap in the middle of the distribution is likely due to the inclusion of QS profiles within both the PF and AR data sets. This is understandable since the IRIS spectrogram can cover a FOV of $130\times175~\text{arcsec}^2$, while an ordinary active region only occupies at most half that FOV, so that all three data sets (AR, SS, PF) are not entirely clean, but rather contain a certain number of mislabeled profiles that cannot be untangled. The overlap towards the tails of the distributions however, indicate the degree to which a non-PF region can produce a PF like profile, and visa versa. For the PF/AR case in Figure \ref{roc_obs_split}, we see that the probability of producing successively more convincing profiles from the opposite class steadily decreases.

\subsection{Spatial and temporal experiments}
The metric scores and  ROC curves for the spatial experiment indicate that the model has learned to correlate PF spectra with higher levels of activity, as seen in the corresponding 1400 SJIs. This activity can be seen towards the top of the SJI in panel B of Figure \ref{iris_ims}. At 80'' away from these visible brightenings, the model does not know if the spectra are from PF or AR datasets, however, the scores and AUCs drastically increase when we present the model with spectra from successively closer regions. This equates to an $f_1$ distribution that starts out centered at 0, and migrates towards 1. Since the metric scores and curves go from random guess performance to perfect classification, fears related to possible biases introduced by yearly varying baseline UV emissions can be dispelled. Furthermore, these results fortify the assumption that mislabeled QS profiles are responsible for the central overlap seen in the probability distribution plot in Figure \ref{roc_obs_split}. This conclusion comes from the fact that at 80 arcseconds off of an active region (where one would expect QS spectra), the model produces metric scores that are entirely consistent with a random guess. This is precisely the score given in the center of the probability distribution plot. With this in mind, the entire enterprise of the spectral flare prediction would greatly benefit from a careful pruning of QS profiles, in a non-full FOV/targeted implementation.

The scores for the spatial experiment do not predict the future location of the flare ribbon which propagates directly over the slit at flare maximum, as seen in Panel D of Figure \ref{iris_ims}. This leads us to believe that the model is simply keeping track of the number of energetic events per time sample.\\ 

The metric scores for the temporal experiment do not vary as extensively, however, the scores and ROC curves indicate that the $f_1$ distribution shifts closer to 1 at about 35 minutes before flare onset, and continues to do so for successive future time steps. One can see a significant increase in metric scores in Figure \ref{space_time_metrics} between t=-60 and t=-50. This increase occurs in conjunction with the first signs of intensity enhancements over the slit, displayed in panel A of Figure \ref{iris_ims}. Each successive step forward in time after the 30 minute mark produces higher metric scores and larger areas under the curve. In other words, the spectra become more PF like in nature the closer we move to flare onset, and thus easier for the model to correctly label. There are a number of reasons why long range temporal tests like these are rare. Firstly, there are a limited number of large flare observations with the IRIS slit positioned directly over the active region that is about to flare. Secondly, this set of observations contains only a small subset where IRIS was recording well in advance of flare onset. Thirdly, the remaining long range PF observations have to be split into training and test sets, which further diminishes this rarefied pool. Lastly, energetic events occurring sufficiently before a flare spoil certain observations. This is indeed the case for this particular flare, although IRIS was recording for a long time prior to flare onset, a C1.0-class flare spoils the possibility of extending the test further back in time. This may also account for the high metric scores observed even 80 minutes before flare onset.\\

For consistency we performed a temporal test on a non-flaring active region. The results in the last panels of \Cref{space_time_metrics,roc_final} demonstrate how the model consistently labels the region as non-flaring, with metric scores below 0.5 indicating the prediction of AR profiles. ROC curves forming on the opposite side of the dividing random choice line indicate the prediction of non-flare profiles, with larger areas under the curve being associated with larger confidences. The scores deviate only slightly with time, in spite of visible small scale energy releases.

\section{Conclusion}\label{Conclusion}
We set out to discover whether spectral profiles collected from QS, SS, AR and PF regions could be distinguished from one another. Each profile was described in terms of 10 features connected to known observables. The distributions of each feature over each of the four solar regions show noticeable differences. These differences can be efficiently leveraged by NNs to build non-linear models that can tell with a high degree of accuracy, whether a profile came from a PF or non-PF region. We exploit the assumption that profiles become increasingly PF like the closer we are to flare onset. Under this reasonable assumption, the models confidence in its classifications can be used as a pre-flare signal, allowing us to build a rudimentary real-time flare forecaster which provides a continual flare probability with each spectral readout. Our findings are as follows:
\begin{enumerate}
\item Results harmoniously confirm the separation between spectra from different solar regions, whether we consider each region's feature distributions, low-dimensional embeddings, or utilize different non-linear classifiers. PF profiles can easily be distinguished from QS profiles, however, this distinction becomes increasingly faded when considering SS and especially AR spectra.
\item The most important features for distinguishing PF from AR spectra to a first order approximation are: triplet emission, followed by intensity, total continuum, line width, k/h ratio, peak separation, k3-height, peak ratios, line asymmetry, and finally, line center position.
\item More intense spectra are often accompanied by triplet emission and decreased total continuum, and have a tendency to form under optically thinner conditions than their QS counterparts. Additionally, deeper reversals are often associated with broader profiles and velocity descriptive features are weak discriminants for the binary PF/AR problem.
\item A light weight 2-hidden layer NN manages to convincingly distinguish between unseen PF and AR spectra, with a TSS score of .57, precision of .78, F1 score of .77, recall of .77, accuracy of .80 and an AUC of .86. This degree of success warrants the conclusion that the same solar regions generate similar spectra, which is consistent with our previous findings \citep{Panos}. 
\item Our trained model can locate the position of pre-flare active regions on the solar disk with a high degree of precision, while the model's temporal performance indicates the utility of \ion{Mg}{2} solar spectra for flare prediction, with TSS scores monotonically increasing from .50 to .78 within a 30 minute margin of flare onset.
\end{enumerate}
This paper demonstrates that spectral data should be considered as an additional source of information for the flare prediction problem, alongside photospheric magnetograms. The importance of incorporating UV data into our predictive algorithms has been eluded to by \cite{UV_Brightening} and further substantiated by us on a finer scale. We have identified particular line features (not all necessarily related to intensity), which carry the capacity for flare prediction on a time scale in accordance with the current research expectations.

\section{Outlook}\label{Outlook}
We offer several ways in which these results could be improved and extended: Removing QS spectra from AR and PF region observations by means of a data count threshold or manually restricting the FOV on a per-observation basis. Early attempts using the reconstruction error of \textit{variational autoencoders} trained on the QS data set have proved a promising direction for cleaning the remaining data sets in a nested fashion. Additional observations would improve the reliability of predictive claims. Repeating the experiment without the intensity feature would be an informative exercise, since this feature may be responsible for overfitting. The interpretability and perhaps the performance of the results could be improved by taking a tighter integration bound around the line cores seen in Figure \ref{profile}. The current broad integration range of $2 \text{\AA}$ may result in a degeneracy and mixing of features such as total continuum and line width. We encourage the exploration of spectra based sequence modeling, incorporating the successes of \cite{LSTM_HMI} by folding in the temporal domain via the application of either simple \textit{Markov models}, \textit{Recurrent Neural Networks} or LSTMs. We have attempted an application of the latter without success. The highly heterogeneous nature of IRIS data, with its multiple operational modes leads to a large spread in cadence which must be reconciled by either incorporating a \textit{masking layer} or by manual interpolation onto a uniform time grid. We intend to examine the thermodynamics associated with high scoring PF spectra using the $\text{IRIS}^2$ database \citep{Alberto}. This could possibly lead to a deeper understanding of the flare triggering mechanism and appears to be the logical next step for extracting an interpretability of our results. We envision that flare prediction models of the future will incorporate multiple streams of diverse time ordered data sets in order to improve on current benchmarks. We hope that the community begins to work towards real-time flare predictions before the next solar maximum.\\

We would like to thank C\'edric Huwyler, S\"{a}m Krucker, Martin Melchior, Sviatoslav Voloshynovskiy and Denis Ullmann for their many fruitful discussions. We used Scikit-Learn modules for the PCA and t-SNE dimensionality reduction as well as the SVM implementation \citep{SK}. All NNs were constructed within the KERAS API \citep{Ker}, while the preprocessing was done in IRISreader, a library specifically developed for handling large volumes of IRIS data \citep{Ced}. We would like to thank the Swiss National Science Foundation for funding this research under grant number 407540\_167158, as well as LMSAL and NASA for allowing us to download all the IRIS data from their servers. IRIS is a NASA small explorer mission developed and operated by LMSAL with mission operations executed at NASA Ames Research center and major contributions to downlink communications funded by ESA and the Norwegian Space Centre.

\bibliographystyle{apj}
\bibliography{journals,references}

\clearpage
\appendix
\section{The bare bones of t-SNE}\label{The bare bones of t-SNE}
The t-SNE algorithm rests on three concepts: 1) statistical distance, 2) information entropy and 3) the Kullback-Leibler (KL) divergence. Concerning the first concept, the pairwise Euclidean distances between each point are converted into statistical distances or similarities $p_{ij} = (p_{j|i}+p_{i|j})/2n$, where the conditional probability $p_{j|i}$ is given by
 \begin{equation}
p _ { j | i } = \frac { \exp \left( - \left\| x _ { i } - x _ { j } \right\| ^ { 2 } / 2 \sigma _ { i } ^ { 2 } \right) } { \sum _ { k \neq i } \exp \left( - \left\| x _ { i } - x _ { k } \right\| ^ { 2 } / 2 \sigma _ { i } ^ { 2 } \right) }.
\end{equation}
Simply stated, t-SNE places Gaussians $g_i$ over each point, where the $i$'th Gaussians is centered at $x_i$ and has a variance $\sigma_i$. These Gaussians impose probability distributions $P_i$ over the other data points, such that the similarity between two points $i$ and $j$ is proportional to the overlap of their Gaussian distributions $g_i$ and $g_j$. It would appear that the whole trick now is figuring out how to select each $\sigma_i$. This leads naturally to the second idea of information entropy or Shannon entropy, defined as
\begin{equation}
S(P_i) = -\sum_jp_{j|i}\text{log}_2p_{j|i}.
\end{equation}
Analogous to the thermodynamic concept of entropy, statistical entropy represents the amount of disorder or uncertainty in a system. If we select a large $\sigma_i$, then $g_i$ spreads over all points such that each point $x_j$ has a similar conditional probability density $p_{j|i}$ under the curve. Because each point has a similar chance of being randomly selected, the system is in a high state of uncertainty or entropy. In contrast, if we select a small $\sigma_i$, then most of the points have conditional probability densities of no significance, finding themselves at the tails of the distribution, while the direct neighbors of $x_i$ carry most of the probability, consequently giving the system a low entropy. In t-SNE, we fix the information entropy by selecting a single scalar value called the perplexity
\begin{equation} 
\alpha = 2^{S(P_i)},
\end{equation}
which in turn fixes all the $\sigma_i$'s in accordance to the underlying distribution of the data set. The imposed probability distributions $P_i$ can then be used to calculate a similarity matrix or distribution $P$, with each entry $p_{ij}$ being the similarity between points $x_i$ and $x_j$. It is important to note that the $\sigma$'s adjust themselves according to the local distribution of data and therefore each $\sigma_i$ can be different. It is now the algorithm's task to represent a noisy version of this similarity distribution within the limits of a 2-dimensional space. Instead of using Gaussian distributions to define a sense of distance, the low-dimensional space uses a Student's t-distribution with a single degree of freedom 
\begin{equation}
p^\dagger _ { i j } = \frac { \left( 1 + \left\| y _ { i } - y _ { j } \right\| ^ { 2 } \right) ^ { - 1 } } { \sum _ { k \neq l } \left( 1 + \left\| y _ { k } - y _ { l } \right\| ^ { 2 } \right) ^ { - 1 } },
\end{equation}
whose heavy tails help negate the problem of \textit{overcrowding}. We now randomly position our points at locations $y_i$ in the low-dimensional space, and with the use of the Student's t-distribution, approximate the actual similarity matrix $P$ with a noisy similarity matrix $P^\dagger$. Finally, the KL-divergence
\begin{equation} 
K L ( P \| P^\dagger ) = \sum _ { i } \sum _ { j } p _ { i j } \log \frac { p _ { i j } } { p^\dagger _ { i j } },
\end{equation}
measures how dissimilar two distributions are. The goal is to minimize the KL-divergence so that the two similarity matrices $P$ and $P^\dagger$ are as close to one another as possible. To this end, we can take the gradient of the KL-divergence
\begin{equation}
\frac{\delta (KL)}{\delta y_{i}}=4 \sum_{j}\left(p_{i j}-p^\dagger_{i j}\right)\left(y_{i}-y_{j}\right)\left(1+\left\|y_{i}-y_{j}\right\|^{2}\right)^{-1},
\end{equation}
and perform gradient descent by nudging each point in the low-dimensional representation such that we approach the minima of the above function. Notice that in order to perform gradient descent, we need a continuously differentiable function. This is in part why the concept of information entropy was introduced, since it can be interpreted as a smooth measure of the effective number of neighbors, that is $\partial\sigma_i/\partial P_i$ exists for each point and over the entire domain. Since the space is not ordinarily convex, the algorithm is not guaranteed to converge to a global minimum, however t-SNE employs a number of caveats to speed up gradient descent and circumvent local minima.

\begin{table}[h]
\caption{Non-PF observations} \centering \small \label{Active region list}
\begin{tabular}{llllll}
\toprule\toprule
\# &  ~~~Date and time &  ~~~~Observation mode   & CAD  & FOV center & ~~OBSID \\ 
& when raster started& &(sec)&~~(arcsec)&\\ \midrule
&&~~~~~~~~~~~~~~~~~~~~~~AR&&\\\midrule
\textcolor{black}{1} &  \textcolor{black}{2015-05-18T14:39} & \textcolor{black}{Large coarse 4-step raster} & ~~\textcolor{black}{21}  & \textcolor{black}{(300,-98)} & \textcolor{black}{3860256971} \\ 
2 &  2015-05-18T16:14 & Large coarse 4-step raster & ~~21  & (315,-95) & 3860256971 \\ 
3 &  2015-05-21T18:59 & Very large sit-and-stare & ~~5  & (-382,398) & 3800507454 \\ 
4 &  2015-07-03T16:59 & Large sparse 8-step raster & ~~45  & (-186,2013) & 3620006130 \\ 
5 &  2015-07-04T10:09 & Very large sit-and-stare & ~~9  & (86,174) & 3860108354 \\ 
6 &  2015-07-04T16:59 & Large sparse 8-step raster & ~~44  & (20,202) & 3620006130 \\ 
7 &  2015-07-28T15:18 &Medium coarse 4-step raster & ~~37  & (-227,-289) & 3660109122 \\ 
8 &  2015-08-07T22:14 & Large coarse 8-step raster & ~~74  & (547,125) & 3860259180 \\ 
\textcolor{black}{9} & \textcolor{black}{2015-08-09T06:15} & \textcolor{black}{Large coarse 8-step raster} & ~~\textcolor{black}{75}  & \textcolor{black}{(-236,-370)} & \textcolor{black}{3860009180} \\ 
10 & 2015-09-16T18:17 & Medium coarse 16-step raster & ~~34  & (-564,-356) & 3600101141 \\ 
11 & 2015-10-17T00:31 & Large sit-and-stare & ~~3  & (-558,-233) & 3660105403 \\ 
\textcolor{black}{12} & \textcolor{black}{2015-07-24T05:35} & \textcolor{black}{Large sit-and-stare} & ~~\textcolor{black}{9}  & \textcolor{black}{(557,-204)} & \textcolor{black}{3620109103} \\
\textcolor{black}{13} & \textcolor{black}{2015-04-08T04:57} & \textcolor{black}{Large sit-and-stare} & ~~\textcolor{black}{5} & \textcolor{black}{(45,-118)} & \textcolor{black}{3860107054}\\
14 & 2015-01-30T11:27 & Very large dense 4-step raster & ~~21  & (-756,161) & 3860607366 \\
15 & 2014-03-29T20:14 & Medium sit-and-stare & ~~17  & (687,-166) & 3820011652 \\
16 & 2014-12-01T15:44 & Large sit-and-stare & ~~10  & (-80,-329) & 3800008053 \\
17 & 2014-03-13T09:35 & Large sit-and-stare & ~~9  & (521,23) & 3820109554 \\
18 & 2014-11-28T21:05 & Very large sit-and-stare & ~~10  & (-34,-322) & 3860009154 \\
\toprule
&& ~~~~~~~~~~~~~~~~~~~~~~SS&&\\\midrule
1 &  2014-07-29T18:00 & Medium sit-and-stare & ~~6  & (359,-49) & 3820007152 \\ 
2 &  2014-07-29T19:59 & Medium dense 16-step raster & ~~56  & (368,47) & 3820005182 \\ 
3 &  2014-07-31T20:07 & Medium dense 16-step raster & ~~88  & (-373,-227) & 3820006082 \\ 
4 &  2014-08-11T11:48 & large sit-and-stare & ~~17  & (-19,122) & 3800011454 \\ 
5 &  2014-10-03T08:08 & Medium dense 4-step raster & ~~21  & (-216,-270) & 3820257165 \\ 
6 &  2015-04-14T19:46 & Medium dense 8-step raster & ~~25  & (-455,321) & 3800104074 \\ 
7 & 2015-04-14T22:58 & Medium dense 8-step raster & ~~25  & (-510,229) & 3800104074 \\ 
8 & 2015-04-15T00:17 & Medium dense 8-step raster & ~~25  & (-557,213) & 3800104074 \\ 
9 & 2015-07-03T11:14 & Large sit-and-stare & ~~32  & (-73,216) & 3880012053 \\ 
10 & 2015-07-16T17:09 & Medium dense 16-step raster & ~~88  & (544,-331) & 3620006035 \\ 
11 & 2015-07-25T16:00 & Large sit-and-stare & ~~9  & (-268,-318) & 3620258103 \\ 
12 & 2015-07-25T18:04 & Large coarse 4-step raster & ~~21  & (-260,-316) & 3620256123 \\ 
13 & 2015-07-25T20:12 & Large coarse 4-step raster & ~~20  & (-242,-314) & 3620106123 \\ 
\toprule
&& ~~~~~~~~~~~~~~~~~~~~~~QS&&\\\midrule
1 &  2014-02-06T12:44 & Large sit-and-stare & ~~5  & (10,32) & 3803257203 \\ 
2 &  2014-02-07T11:29 & Large sit-and-stare & ~~5  & (236,28) & 3803257203 \\ 
3 &  2014-02-08T13:32 & Large sit-and-stare & ~~5  & (503,29) & 3803257203 \\
4 &  2014-02-11T05:10 & Large sit-and-stare & ~~4  & (19,40) & 3864255653 \\
5 &  2014-05-11T06:49 & Medium sparse 2-step raster & ~~19  & (-89,-172) & 3800258458 \\ 
6 &  2014-05-15T14:09 & large sit-and-stare & ~~5  & (11,-2) & 3820357403 \\ 
7 &  2014-05-16T07:58 & Medium sparse 2-step raster & ~~19  & (-3,-3) & 3800258458 \\
8 &  2014-06-26T22:32 & Large dense 16-step raster & ~~56  & (362,190) & 3820005183 \\
9 & 2014-06-28T16:48 & Medium dense 16-step raster & ~~50  & (-178,121) & 3820505482 \\
10 & 2014-06-28T21:42 & Medium dense 16-step raster & ~~50  & (234,-180) & 3820505482 \\
11 & 2014-09-02T07:30 & Medium dense 4-step raster & ~~37  & (522,-91) & 3800258465 \\
12 & 2014-10-07T07:54 & Medium sparse 2-step raster & ~~19  & (185,79) & 3800258458 \\
13 & 2014-12-14T15:38 & Large coarse 8-step raster & ~~41  & (337,248) & 3800106080 \\ 
14 & 2015-10-10T23:34 & Medium dense 16-step raster & ~~55  & (-26,2) & 3620005935 \\ 
15 & 2014-09-15T22:09 & Very large dense 4-step raster & ~~21  & (-226,375) & 3820107266 \\ 
16 & 2015-02-18T20:01 & Very large dense 4-step raster & ~~62  & (820,61) & 3800010066 \\ 
17 & 2014-02-06T12:44 &  Large sit-and-stare & ~~5  & (5,31) & 3803257203 \\ 
18 & 2014-12-14T17:15 &  Large coarse 8-step raster & ~~42  & (351,248) & 3800106080 \\ 
 \bottomrule
\label{qsssar}
\end{tabular}
\end{table}

\begin{table}[h]
\caption{PF observations} \centering \small \label{Pre-flare list}
\begin{tabular}{lllllll}
\toprule\toprule
\# &  Class  &  ~~~Date and time &  ~~~~~~~Observation mode   & CAD   &FOV center & ~~OBSID \\ 
& & when raster started& & (sec) &~~(arcsec)&\\\midrule
1&  M1.0&  2014-06-12T18:44& Medium coarse 8-step raster&  21 & (-670,-306)& 3863605329 \\ 
2& M1.3& 2014-10-26T18:52& Large sit-and-stare& 16 & (648,-287)& 3864111353 \\
3& M1.0 &2014-11-07T09:37& Large coarse 16-step raster& 23 & (-646,224)&  3860602088 \\
4& M1.1& 2014-09-06T11:23& Large sit-and-stare& 9 & (-709,-298)& 3820259253 \\
5 &  M1.1&  2015-08-21T16:01& Medium dense 32-step raster& 102 & (-467,-336)& 3660104044\\
6& M1.4& 2015-03-12T05:45& Large sit-and-stare& 5 & (-185,-190)& 3860107053 \\
7& M1.8&2014-02-12T21:50&Large coarse 8-step raster&42 &(140,-90)&3860257280\\
8& M1.8&2015-03-11T04:46&Large coarse 8-step raster&75 &(-430,-194)&3860259280\\
9& M2.3&2014-11-09T15:17&Large coarse 4-step raster&37 &(-217,-205)&3860258971\\
10& M3.4&2014-10-27T20:56&Large sit-and-stare&16 &(779,-271)&3864111353\\
\textcolor{black}{11}& \textcolor{black}{M3.9}&\textcolor{black}{2014-06-11T18:19}&\textcolor{black}{Medium coarse 8-step raster}&\textcolor{black}{21} &\textcolor{black}{(-781,-306)}&\textcolor{black}{3863605329}\\
12& M6.5&2015-06-22T17:00&Large sparse 16-step raster&33 &(72,192)&3660100039\\
13& M8.7&2014-10-21T18:10&Large sit-and-stare&16 &(-359,-316)&3860261353\\
14& X1.0&2014-10-25T14:58&Large sit-and-stare&5 &(408,-319)&3880106953\\
15& X1.0&2014-03-29T14:09&Very large coarse 8-step raster&72 &(490,282)&3860258481\\
\textcolor{black}{16}& \textcolor{black}{X1.6}&\textcolor{black}{2014-09-10T11:28}&\textcolor{black}{Large sit-and-stare}&\textcolor{black}{9} &\textcolor{black}{(-137,125)}&\textcolor{black}{3860259453}\\
17& X1.6&2014-10-22T08:18&Very large coarse 8-step raster&131 &(-292,-303)&3860261381\\
\textcolor{black}{18}& \textcolor{black}{X2.0}&\textcolor{black}{2014-10-27T14:04}&\textcolor{black}{Large coarse 8-step raster}&\textcolor{black}{26} &\textcolor{black}{(727,-299)}&\textcolor{black}{3860354980}\\
\textcolor{black}{19}& \textcolor{black}{X2.1}&\textcolor{black}{2015-03-11T15:19}&\textcolor{black}{Large coarse 4-step raster}&\textcolor{black}{16} &\textcolor{black}{(-353,-197)}&\textcolor{black}{3860107071}\\
 \bottomrule
\label{pf}
\end{tabular}
\end{table}

\end{document}